\documentclass[a4paper,final]{article}         
\usepackage{a4wide}
\usepackage{natbib}
\usepackage[latin1]{inputenc}

\usepackage{amsmath}
\usepackage{amssymb}
\usepackage{graphicx}
\usepackage{url}
\usepackage{xcolor}
\usepackage[draft,silent]{fixme}

\usepackage{lscape}

\usepackage[T1]{fontenc}
\usepackage[sc]{mathpazo}
\def\PE{\mathbb{E}}
\def\PP{\mathbb{P}}
\def \1{\mathbf{1}}
\def\inproba{\stackrel{p}{\longrightarrow}}
\def\inlaw{\stackrel{d}{\longrightarrow}}
\newcommand{\inv}[1]{\frac{1}{#1}}

\newcommand{\iid}{\stackrel{iid}{\sim}}

\usepackage[bookmarksnumbered,colorlinks, bookmarks, breaklinks, linktocpage]{hyperref}
\newtheorem{theorem}{Theorem}
\newtheorem{remark}{Remark}

\title{Distributed detection/localization of change-points in
  high-dimensional network traffic data\footnote{the original publication is available at \url{www.springerlink.com} or \url{http://dx.doi.org/10.1007/s11222-011-9240-5}}}
\author{
Alexandre Lung-Yut-Fong, Céline Lévy-Leduc, Olivier Cappé
}
\date{22/02/2011}
\begin{document}
\maketitle

\begin{abstract}
  We propose a novel approach for distributed statistical detection of change-points in high-volume
  network traffic. We consider more specifically the task of detecting and identifying the targets
  of Distributed Denial of Service (DDoS) attacks.
  The proposed algorithm, called \emph{DTopRank}, performs distributed network anomaly detection by
  aggregating the partial information gathered in a set of network monitors. In order to address
  massive data while limiting the communication overhead within the network, the approach combines
  record filtering at the monitor level and a nonparametric rank test for doubly censored time
  series at the central decision site.
  The performance of the \emph{DTopRank} algorithm is illustrated both on synthetic data as well as
  from a traffic trace provided by a major Internet service provider.

\end{abstract}

\section{Introduction}

Detecting malevolent behaviors has become a prevalent concern for the security of network
infrastructures, as exemplified by the, now common, attacks against major web services
providers. In this contribution, we consider more specifically the case of DDoS (Distributed Denial
of Service) type of attacks where many different sources transmit data over the network to a few
targets so as to flood resources and, eventually, cause disruptions in service.

Several methods for dealing with DDoS attacks
have been proposed. They can be arranged into two categories: signature-based approaches
and statistical methods.
The former operate by comparing the observed patterns
of network traffic with known attack templates. Obviously,
this methodology only applies for detecting anomalies that
have already been encountered and characterized. The second
type of approaches relies on the statistical analysis of network patterns and can thus
potentially detect any type of network anomalies. The basic statistical modelling for this task is to assume that network anomalies lead to abrupt changes
in some network characteristics.
Hence, most
statistical methods for detection of network anomalies are cast in the framework of statistical change-point detection, which is a familiar topic in statistics, see, e.g., \cite{Basseville:1993,Brodsky:1993,Csorgo:1997}, and references therein.

Two different approaches to change-point detection are usually distinguished:
the detection can be retrospective and hence with a fixed 
delay (batch approach) or online, with a minimal average delay (sequential
approach). In the field of network security, a widely used
change-point detection technique is the cumulated sum (CUSUM)
algorithm described in \cite{Basseville:1993} which is a sequential
approach. It has, for instance, been used by \cite{Wang:2002} 
and by \cite{Siris:2004} for detecting DoS attacks of the TCP 
(Transmission Control Protocol) SYN flooding type.
This attack consists in exploiting the TCP three-way hand-shake mechanism and its limitation in maintaining half-open
connections. More precisely, when a server receives a SYN packet, it returns a SYN/ACK packet to the client. Until 
the SYN/ACK packet is acknowledged by the client, the connection remains half-opened for a period
of at most the TCP connection timeout. A backlog queue is built up in the system memory of the server
to maintain all half-open connections, this leading to a saturation of the server. In 
\cite{Siris:2004}, the authors use the CUSUM algorithm to detect a change-point in
 the time series corresponding to the aggregation of the SYN
 packets received by all the requested destination IP
addresses. 
With such an approach, it is only possible to set off an alarm when a massive change occurs in the aggregated
series; it is moreover impossible to identify the attacked IP addresses.

Given the nature of a TCP/SYN flooding attack, the attacked IP addresses may be identified by applying multiple change-point 
detection tests, considering each of the time series formed by counting the 
number of TCP/SYN packets received by individual IP addresses.
This idea is used in \cite{Tartakovsky:2005} where
a multichannel detection procedure, which is a refined
version of the previously described algorithm, is proposed: 
it makes it possible to detect changes which occur in a
channel and which could be obscured by the normal traffic in the other channels 
if global statistics were used.

When analyzing wide-area-network traffic, however, it is not possible anymore to consider
individually all the possible target addresses for computational reasons.
For instance, the data used for the evaluation of the proposed method
(see Section \ref{sec:real}) contain several thousands of
distinct IP addresses in each one-minute time slot.
 In order to detect anomalies in  
such massive data within a reasonable time span, it is impossible to analyze the 
time series of all the IP addresses receiving TCP/SYN packets. That is why dimension reduction
techniques have to be used. Three main approaches have been proposed. The first one uses
Principal Component Analysis (PCA) techniques, see \cite{Lakhina:2004}. 
The second one uses random aggregation (or sketches), see \cite{Krishnamurthy:2003}
and the third one is based on record filtering, see \cite{Levy:Roueff:2009}.
Localization of the anomalies is possible with the second and third approaches but not with the
first one. By localization, we mean finding the attacked IP addresses.

In the approaches mentioned above, all the data are sent to a central analysis site,
called the collector in the sequel, in which a decision is made
concerning the presence of an anomaly. These methods are called
centralized approaches. A limitation of these methods is that 
they are not adapted to large networks with massive data since, in
this case, the communication overhead within the network becomes
significant. The approach that we propose in this paper consists
in processing the data within the network (in local monitors)
in order to send to the collector only the most relevant data.  
These methods are called,
in the sequel, decentralized or distributed approaches.
In \cite{huang:nips}, a method
to decentralize the approach of \cite{Lakhina:2004} is considered but, 
as previously explained, with such a method localizing the network anomaly is impossible.


The main contribution of this paper is an efficient way of decentralizing the \emph{TopRank}
algorithm introduced in \cite{Levy:Roueff:2009}. The proposed algorithm, termed \emph{DTopRank}
(for Distributed \emph{TopRank}), uses the \emph{TopRank} algorithm locally in each monitor and
only sends the most relevant data to the collector. The data sent by the different local monitors
are then aggregated in a specific way that necessitates the development of a novel nonparametric
rank test for doubly censored data that generalizes the proposal of \cite{Gombay:Liu:2000}. The
\emph{DTopRank} algorithms makes it possible to achieve a performance
that is on a par with the fully
centralized \emph{TopRank} algorithm while minimizing the data that needs to be send from the
monitors to the collector.

The paper is organized as follows. In Section \ref{sec:description}, we describe the
\emph{DTopRank} method and determine the limit in distribution of the proposed test statistic under
the null hypothesis that there is no network anomaly. The performance of the proposed algorithm
(implemented in C language) is then assessed both using a real traffic trace provided by a major
Internet Service Provider (Section \ref{sec:real}) as well as on synthetic data (Section
\ref{sec:simul}). In both cases, \emph{DTopRank} is compared both to the centralized \emph{TopRank}
algorithm and to a simpler baseline decentralized algorithm based on the use of the Bonferroni
correction.

\section{Description of the methods} \label{sec:description}

The raw data that is analyzed consists of flow-level
summaries of the communications on the network. These
include, for each data flow, the source and destination IP addresses, 
the start and end time of the communication as well as
the number of exchanged packets. All of this information is contained
in the standard Netflow format.


Depending on the type of anomaly to be detected, one needs to consider specific aspects of the network traffic. In the case of the TCP/SYN flooding, the quantity of interest is the number
of TCP/SYN packets received by each destination IP address per unit of time. We denote by $\left(N_i(t)\right)_{t\geq1}$ the discrete time series formed by counting the number of TCP/SYN packets received by the destination IP address $i$ in the $t$-th sub-interval of size $\Delta$ seconds, where $\Delta$ is the sampling period.

The centralized \emph{TopRank} algorithm analyzes these global packets counts. In our case however,
we consider a monitoring system with a set of local monitors $M_1,\dots,M_K$, which collect and
analyze the locally observed time series. As as consequence of decentralized processing, the
packets sent to a given destination IP address are not observed at all monitors, although some
overlap may exist, depending on the routing matrix and the location of the monitors. We thus denote
by $N_i^k(t)$ the number of TCP/SYN packets transiting to the destination IP address $i$ in the
sub-interval indexed by $t$, as observed by the $k$-th monitor. In the proposed batch approach,
detection is performed from the data observed during an observation window of duration
$P\times\Delta$ seconds. The goal is to detect change-points in the aggregated time series
$\left(N_i(t)\right)_{t\geq1}$ using only the local time series $\left(N_i^k(t)\right)_{t\geq1}$
for each $k\in\{1,\dots,K\}$ and a quantity of data transmitted to the collector that is as small
as possible.

\subsection{The \emph {DTopRank} method}

The \emph{DTopRank} algorithm operates at two distinct levels: the local processing
step within the local monitors $M_1,\dots,M_K$ and the aggregation and global change-point detection step within
the collector.

\subsubsection{Local processing}\label{subsec:loc_proc}

The local processing of \emph{DTopRank} consists of the four steps
described below, which are applied in each of the $K$ monitors. The
first three steps are similar to the \emph{TopRank} algorithm applied
to the local series of counts $\left(N_i^k(t)\right)_{1\leq t\leq
  P}$. The second and third steps are however modified by introducing
a lower censoring value for each analyzed series so as to make possible global aggregation at the collector level. In this section, the superscript $k$, corresponding to the monitor index, is dropped to alleviate the notations.

\paragraph{1. Record filtering: }

For each time index $t\in\{1,\dots,P\}$, the indices of the $M$ largest counts $N_i(t)$ are
recorded and labeled as $i_1(t),\dots,i_M(t)$ to ensure that $N_{i_1(t)}(t)\geq
N_{i_2(t)}(t)\geq \dots \geq N_{i_M(t)}(t)$. In the sequel, $\mathcal{T}_{M}(t)$ denotes the set
$\{i_1(t),\dots,i_M(t)\}.$ We stress that, in order to perform the following steps, we only need to
store the variables $\{N_i(t),\,i\in\mathcal{T}_M(t),\,t=1,\dots,P\}$.

\paragraph{2. Creation of censored time series: }

For each index $i$ selected in the previous step ($i\in\bigcup_{t=1}^{P} \mathcal{T}_{M}(t)$),
the censored time series  is built. This
time series is censored since $i$ does not necessarily belong to
the set $\mathcal{T}_M(t)$ for all indices $t$ in the observation window,
in which case, its value $N_i(t)$ is not available and is censored using the
upper bound $N_{i_M(t)}(t)=\min_{i\in\mathcal{T}_M(t)}N_i(t)$. More
formally, the censored time series  $(X_i(t),\delta_i(t))_{1\leq t\leq P}$
are defined, for each $t\in\{1,...,P\}$, by
\begin{align*}
&(X_i(t),\delta_i(t))=\left\{
\begin{array}{cc}
(N_i(t),1), & \textrm{ if } i\in \mathcal{T}_M(t)\\
(\displaystyle\min_{j\in \mathcal{T}_M(t)}N_j(t),0), & \textrm{ otherwise.} 
\end{array} 
\right.
\end{align*}
The value of $\delta_i(t)$ indicates whether the corresponding value
$X_i(t)$ has been censored or not.
Observe that, by definition, $\delta_i(t)=1$ implies that $X_i(t)=N_i(t)$
and 
$\delta_i(t)=0$ implies that $X_i(t)\geq N_i(t).$
%
%
We also define the upper and lower bounds of $X_i(t)$ by $\overline{X}_i(t)=X_i(t)$ and
$\underline{X}_i(t)=X_i(t)\delta_i(t)$, respectively.


In order to process a fixed number $S$ of  time series
instead of all those in $\bigcup_{t=1}^{P} \mathcal{T}_{M}(t)$ 
(at most $M\times P$), we only build the time series
corresponding to the index $i$ in
the list $i_1(1),\dots,i_1(P), i_2(1),\\
\dots,i_2(P)$, $i_3(1),\dots$ where the indices $i_k(t)$ are defined
in the previous step.



\paragraph{3. Change-point detection test: }

In \cite{Levy:Roueff:2009}, the nonparametric 
test proposed by \cite{Gombay:Liu:2000} is used for detecting
change-points in censored data. Here, this test is extended
in order to detect change-points in doubly censored time
series so that the same procedure can be applied both in the local
monitors and within the collector. This test, described hereafter, is applied
to each time series created in the previous stage and the
corresponding $p$-value is computed, a small value suggesting a potential anomaly.


Let us now further describe the statistical test that we perform.
This procedure aims at testing from the observations 
previously built $(\underline{X}_i(t),\overline{X}_i(t))_{1\leq t\leq P}$
if a change occurred in this time series for a given $i$.
More precisely, if we drop the subscript $i$ for convenience in the description
of the test, the tested hypotheses are:

$(H_0)$: ``$(\underline{X}(t),\overline{X}(t))_{1\leq t\leq P}$ 
are independent and identically distributed. '' 

$(H_1)$: ``There exists some $r$ such that
$\left((\underline{X}(1),\overline{X}(1)),\dots,\right.$
$\left.(\underline{X}(r),\overline{X}(r))\right)$ and \\
$\left((\underline{X}(r+1),\overline{X}(r+1)),\dotsc,
(\underline{X}(P),\overline{X}(P))\right)$ have a different distribution. ''

To define the proposed test statistic,
define, for each $s,t$ in $\{1,\dots,P\}$,
$$
h(s,t)=\1(\underline{X}(s)>\overline{X}(t))-\1(\overline{X}(s)<\underline{X}(t))\;,
$$
where $\1(E)=1$ in the event $E$ and 0 in its complementary set,
and
\begin{equation}\label{eq:Y_i}
Y_{s}=\frac{U_{s}}{\sqrt{\sum_{t=1}^{P}U^{2}_{t}}}\; , \quad
\textrm{with} \quad
U_s=\sum_{t=1}^P h(s,t)\; .
\end{equation}
The test statistic is then given by
$$
W_P=\max_{1\leq t\leq P}|\sum_{s=1}^{t}Y_{s}|\; .
$$
The following theorem, which is proved in appendix,
provides, under mild assumptions, the limiting distribution of 
$W_P$, as $P$ tends to infinity, under the null hypothesis
and thus provides a way of computing the $p$-values of the test.

\begin{theorem}\label{theo:null_hyp}
Let $(\underline{X},\overline{X})$ be a $\mathbb{R}^2$-valued random vector
such that
\begin{equation}\label{eq:cond1}
\PP(F(\underline{X}^{-}) + G(\overline{X}) = 1) < 1\; ,
\end{equation}
where $F$ is the c.d.f. of  $\overline{X}$,
$G$ the c.d.f. of $\underline{X}$ and $F(x^-)$ denotes the left
limit of $F$ at point $x$. Let
$(\underline{X}(t),\overline{X}(t))_{1\leq t\leq P}$ be
i.i.d. random vectors having the same distribution as $(\underline{X},\overline{X})$,
then, as $P$ tends to infinity,
\begin{equation}\label{eq:enonceConvBrownian}
  \sup_{0\leq u\leq 1}|\sum_{s=1}^{\lfloor Pu\rfloor}Y_{s}| \inlaw
 B^{\star}:=\sup_{0\leq u\leq 1}|B(u)|\;,
\end{equation}
where $\{B(u)\; , 0\leq u\leq 1\}$ denotes the Brownian Bridge and
$\stackrel{d}{\longrightarrow}$ refers to convergence in distribution.
\end{theorem}

Theorem \ref{theo:null_hyp} of this paper thus extends Theorem 1 of
\cite{Gombay:Liu:2000}, where only one-sided censoring
was considered and continuity of the random variables was assumed.

\begin{remark}\label{rem1}
Theorem \ref{theo:null_hyp} provides a way of controlling the
asymptotic false-alarm rate, for large enough observation sizes.
The only requirement is (\ref{eq:cond1}), which is a minimal condition.
In particular, if the random variables $\underline{X}$
and $\overline{X}$ both have a continuous c.d.f., (\ref{eq:cond1}) holds
whenever $\PP(\underline{X}=\overline{X})>0$, that is, when the
probability of not being censored is positive.
Indeed, 
$\PP(F(\underline{X}) + G(\overline{X})= 1)
=\PP(\{F(\underline{X}) + G(\overline{X})=
1\}\cap\{\overline{X}=\underline{X}\})
+\PP(\{F(\underline{X}) + G(\overline{X})=
1\}\cap\{\overline{X}\neq\underline{X}\})
=\PP(\{2 F(\overline{X})=1\}\cap\{\overline{X}=\underline{X}\})
+\PP(\{F(\underline{X}) + G(\overline{X})=
1\}\cap\{\overline{X}\neq\underline{X}\}).$
Observe that the first probability is smaller than
$\PP(2F(\overline{X})=1).$ Using that $F$ is continuous, 
$F(\overline{X})$ has a uniform distribution on $[0,1]$
and thus $\PP(2F(\overline{X})=1)=0.$ Thus, 
$\PP(F(\underline{X}) + G(\overline{X})=
1)\leq\PP(\overline{X}\neq\underline{X})
=1-\PP(\overline{X}=\underline{X})$.
In practice, the $p$-values deduced from
Theorem \ref{theo:null_hyp} are reliable
whenever the observation size $P$ is large enough and some
non-censored values have indeed been observed.
\end{remark}

\begin{remark}
The i.i.d. assumption in Theorem \ref{theo:null_hyp} may seem surprising
in light of the ubiquity of long-range dependence phenomenons in
aggregated network traffic measurements; see for instance
\cite{park:Hernandez-campos:marron:donelson:2005} and references therein.
However, the assumption here applies to single Origin-Destination flows
which, most often, do not exhibit strong autocorrelations; see
\cite{susitaival:juva:peukuri:2006} for further discussion of this issue.
\end{remark}

\begin{remark}
In practice, the computation of the quantities $(\sum_{s=1}^{t}Y_{s})_{1\leq
  t\leq P}$ can be done in $O(P)$ operations only using the alternate form of $U_s$ in term of the empirical cumulative distribution functions of $\overline{X}(t)$ and $\underline{X}(t)$ (see Eq.~(\ref{eq:Ui_p}) in appendix).
\end{remark}

Based on  (\ref{eq:enonceConvBrownian}),
we take for the change-point detection test the following $p$-value: 
$Pval(W_P)$, where for all positive $b$ (see, for instance,
\citealp{billingsley:1968}, p. 85),
$$
Pval(b)=\mathbb{P}(B^{\star} >b)=2\sum_{j=1}^{\infty} (-1)^{j-1}
e^{-2j^2b^2}\; .
$$
%

\paragraph{4. Selection of the data to be transmitted to the collector: }

We select in each monitor $M_k$ the $d$ censored time series having the
smallest $p$-values and send them to the collector. 
Thus, the collector receives at most $d\times K$ censored
time series, instead of $\sum_{k=1}^{K}D_{k}$, where $D_{k}$ is the number of
destination IP addresses seen by the $k$th monitor, if a centralized approach was used.

\subsubsection{Aggregation and change-point detection test in the collector}

Within the collector, the lower and upper bounds of the aggregated time series 
$(\underline{Z}_i(t),\overline{Z}_i(t))_{1\leq t\leq P}$ associated to
the IP address $i$ are then built as follows:
\begin{equation}\label{eq:aggregation}
\underline{Z}_i(t)=\sum_{k\in\mathcal{K}}\underline{X}_i^{(k)}(t)\ \textrm{  and  }\ 
\overline{Z}_i(t)=\sum_{k\in\mathcal{K}}\overline{X}_i^{(k)}(t)\; ,
\end{equation}
where $(\underline{X}_i^{(k)}(t),\; t=1,\dots,P)$ and
$(\overline{X}_i^{(k)}(t),\;t=1,\dots,P)$ are the time series associated to
the IP address $i$ computed by the monitor $M_k$, and $\mathcal{K}$
is the set of monitors which have transmitted series pertaining to the IP address $i$.
Then, the test described in step 3 of the local processing is applied
to the time series 
$(\underline{Z}_i(t),\;t=1,\dots,P)$ and
$(\overline{Z}_i(t),\;t=1,\dots,P)$.
An IP address $i$ is thus claimed to be attacked at a given false alarm rate 
$\alpha\in(0,1)$, if $Pval(W_P)<\alpha$, and the change-point time is estimated with
$\hat r=\arg\max_{1\leq t\leq P} |\sum_{s=1}^{t}Y_{s}|$.

As noted in Remark~\ref{rem1} above, Theorem \ref{theo:null_hyp} may be safely applied
to the aggregated time series $(\underline{Z}_i(t),\overline{Z}_i(t))$ as long as they are not fully censored. By definition, the aggregated series is more
censored than the individual series $(\underline{X}_i^{(k)}(t), \overline{X}_i^{(k)}(t))$ detected at monitor level.
On the other hand, for a given address $i$, only the series that are
among the $d$ series having the smallest $p$-values at the monitor level are aggregated
at the collector level. Hence, the collector usually aggregates strictly less than $K$ series
and only aggregates potentially significant series. In the
experimental conditions described in Sections~\ref{sec:real}
and~\ref{sec:simul} below ($d$ ranging from one to five, $P=60$ and
$M=10$), the average number of uncensored values in the aggregated
series is 18.7 (out of 60) and the average rate of completely censored time
series is equal to 0.62$\%$. The parameter that has the largest
influence on these values is the depth $M$ of the record filtering
buffer: increasing $M$ reduces censoring 
(for $M=20$, the proportion of uncensored data in the aggregated
series raises up to 31.2 out of 60
and the average rate of completely censored time
series is equal to 0.017$\%$), on the other hand, the required memory and computation time for the record filtering step are both increasing with $M$. Hence, $M=10$ represents a good tradeoff --~see also Section~\ref{sec:simul} for further discussion.

\subsection{The \emph{BTopRank} method}

In the sequel, the \emph{DTopRank}
algorithm is compared with a simpler approach using, instead of the aggregation
step, a simple Bonferroni correction of the $p$-values determined in each monitor.
More precisely, in \emph{BTopRank} an IP address is claimed
to be attacked at the level $\alpha\in (0,1)$ within the collector if
at least one local monitor has computed a
$p$-value smaller than $\alpha/K$, namely if $K(\inf_{1\leq k\leq
  K}\textrm{Pval}_k) <\alpha$, $\textrm{Pval}_k$ being the $p$-value
computed in the monitor $k$.

\section{Application to real data}\label{sec:real}

This section summarizes the results obtained by the \emph{DTopRank} and
\emph{BTopRank} algorithms applied to an actual Internet 
traffic trace provided by a major Internet service provider.

\subsection{Description of the data}

\begin{figure*}
\begin{center}
\begin{tabular}{ccc}
\includegraphics*[width=5cm]{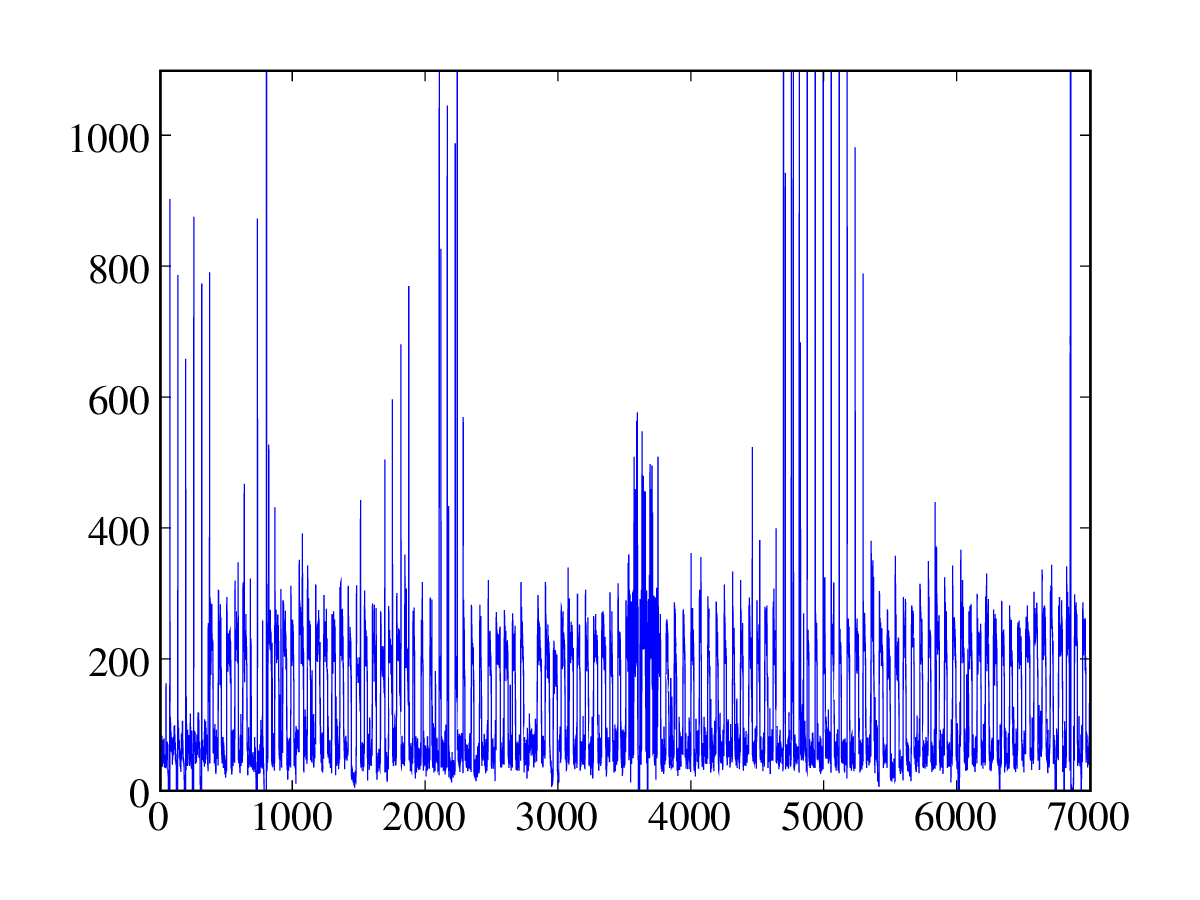}
& \includegraphics*[width=5cm]{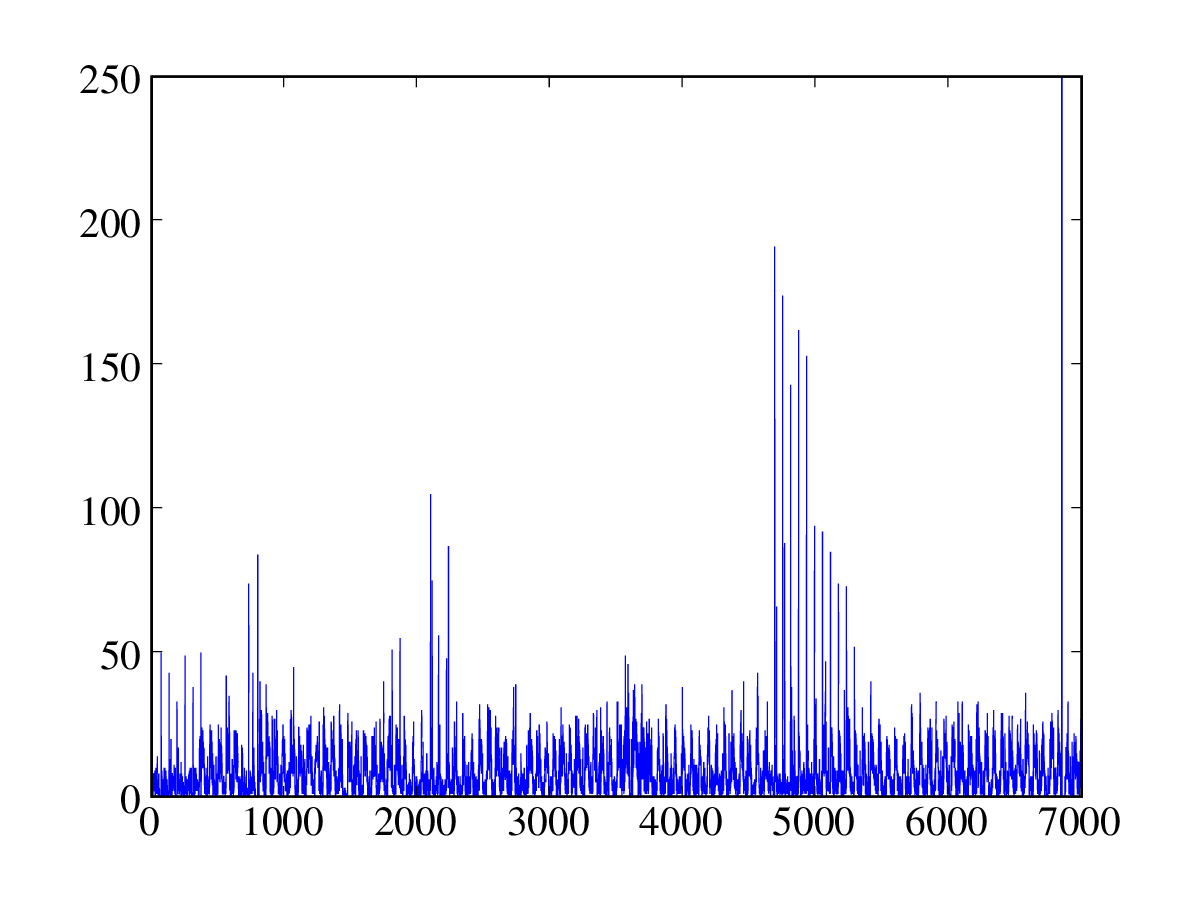}
&  \includegraphics*[width=5cm]{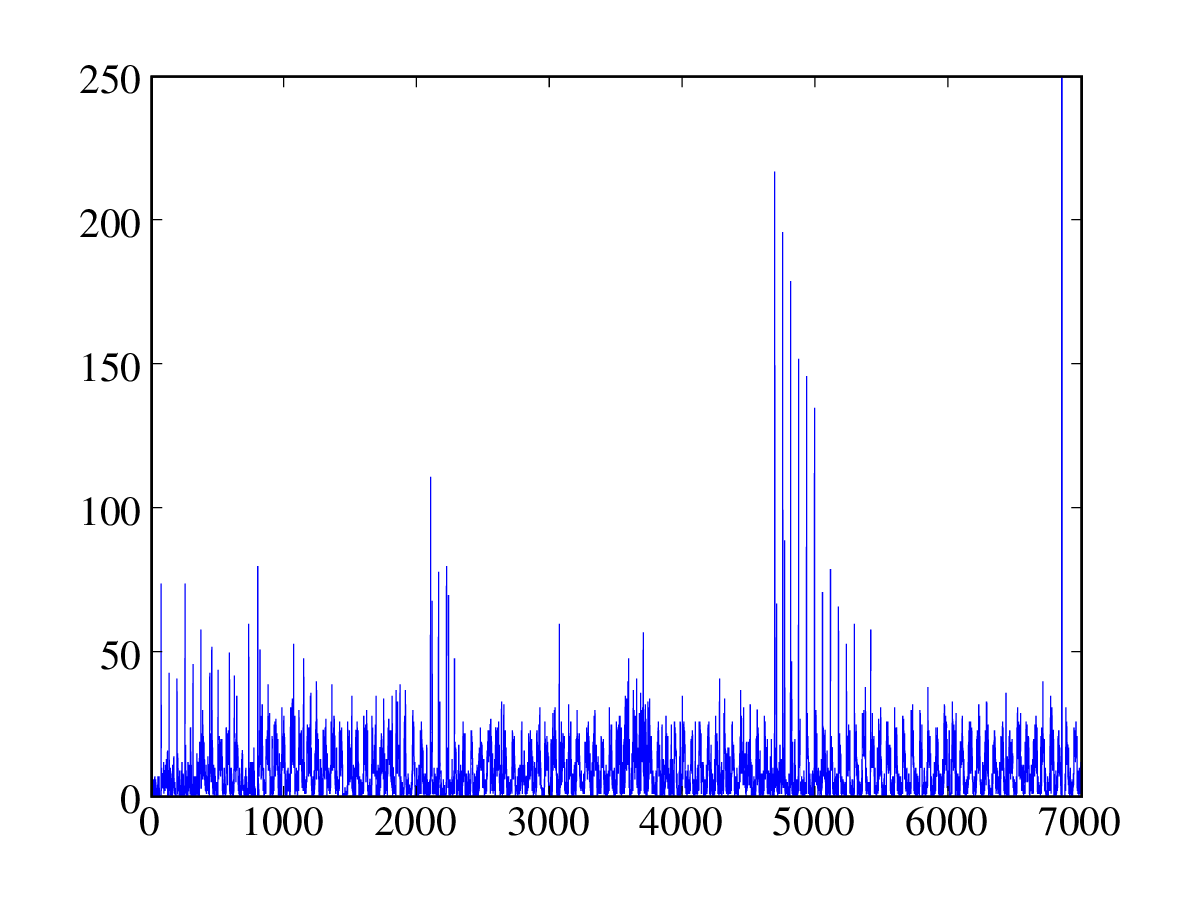}\\
(a) - Global traffic &(b) - Traffic in monitor 1 &(c) - Traffic in monitor 2 \\
\includegraphics*[width=5cm]{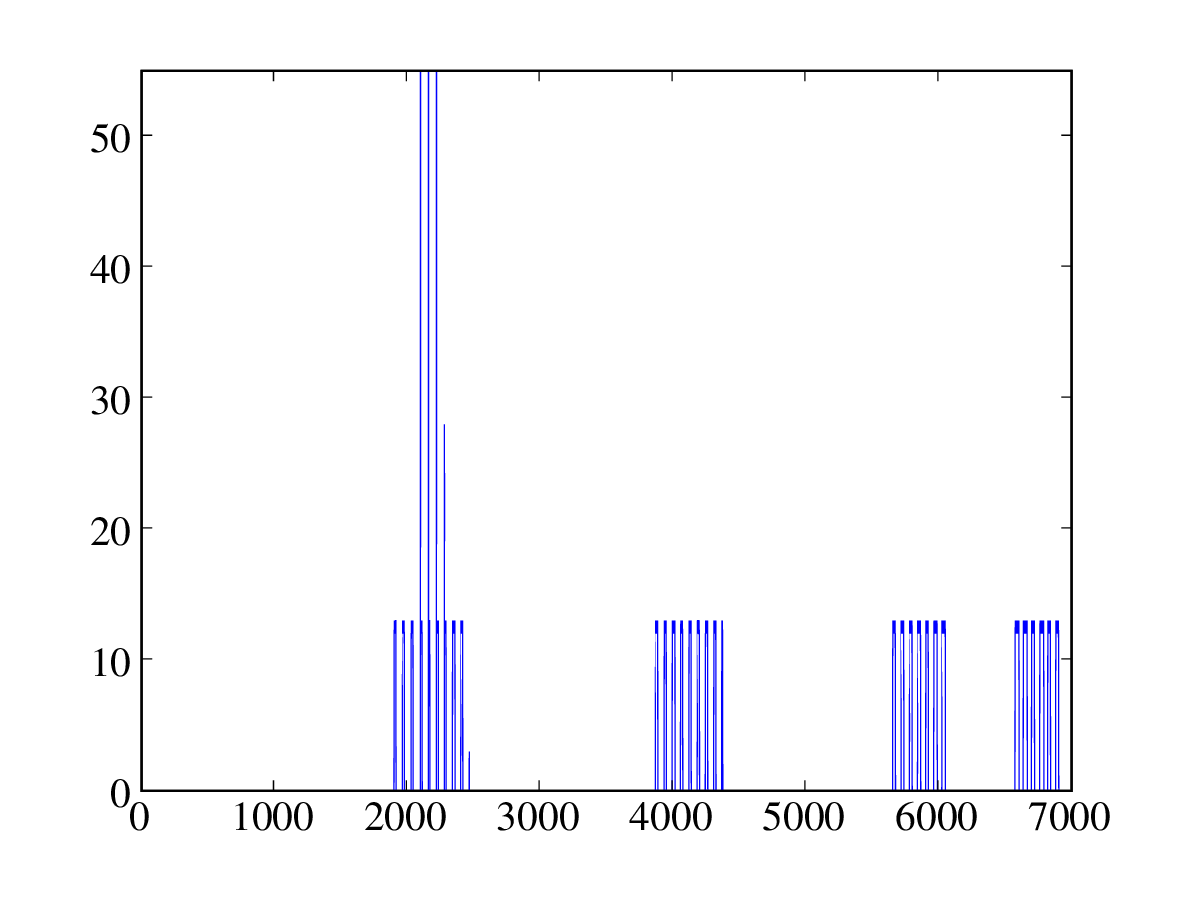}
& \includegraphics*[width=5cm]{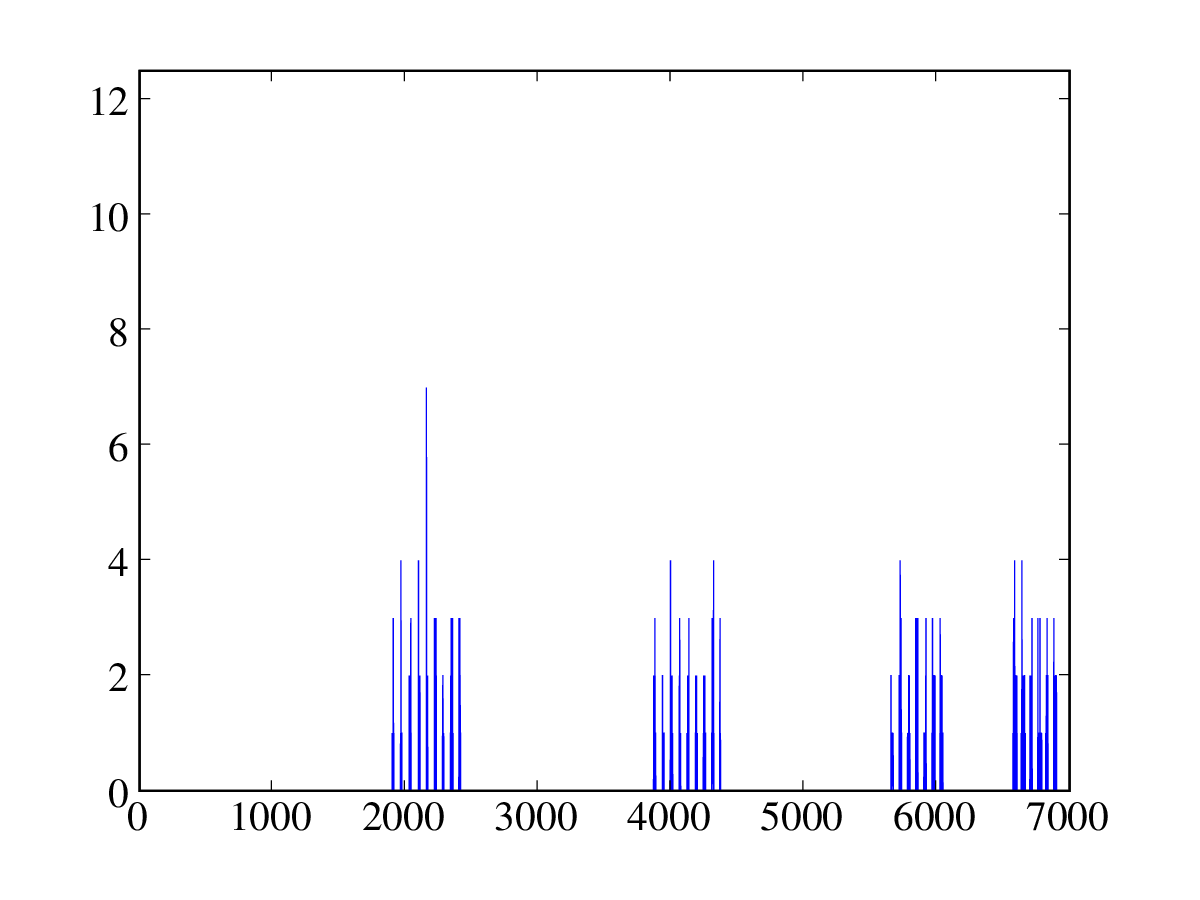}
& \includegraphics*[width=5cm]{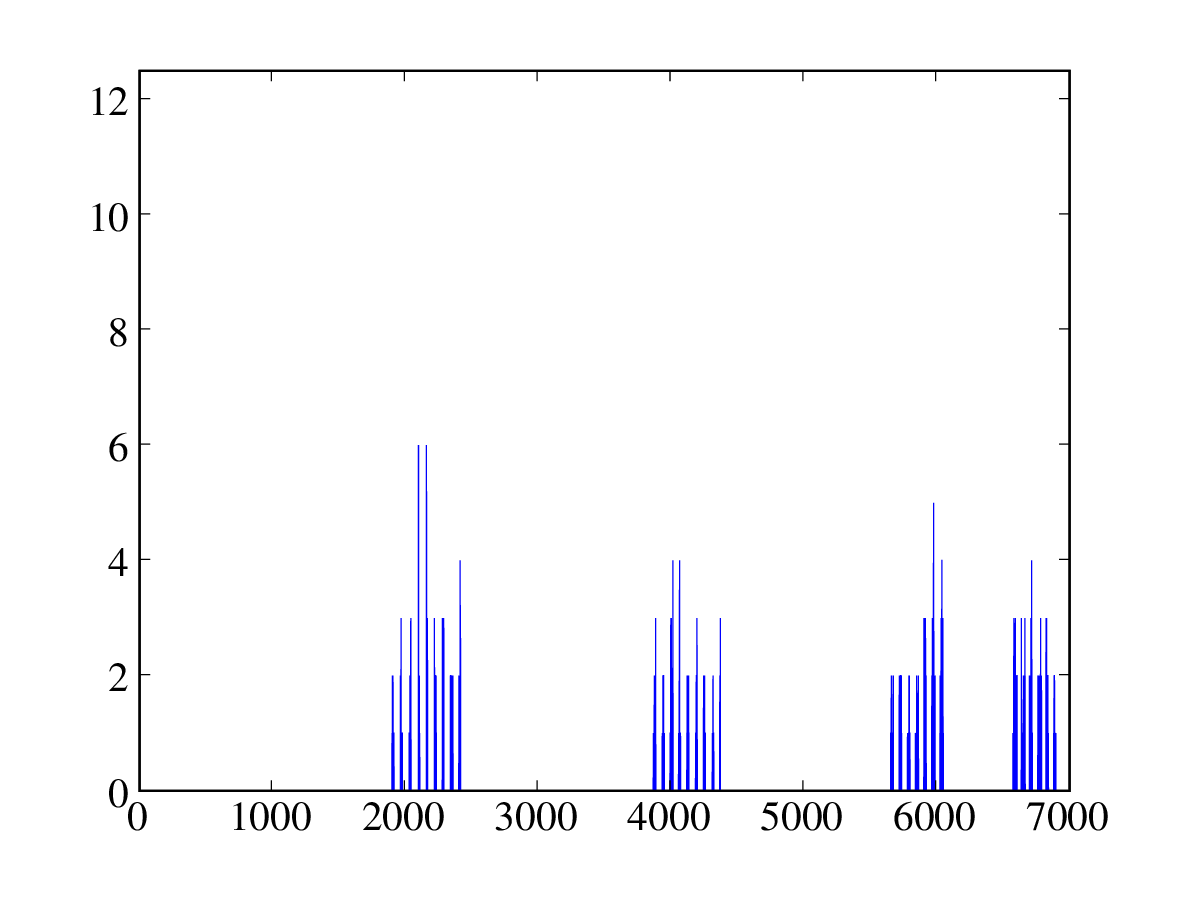}\\
(d) - Attacks &(e) - Attacks in monitor 1 &(f) - Attacks in monitor 2
\end{tabular}
\end{center}
\caption{\footnotesize{Number of TCP/SYN packets globally exchanged (top) and received by the 4 attacked
IP addresses (bottom) in the original data (a,d) and within two particular
monitors (b, c, e, f). Note that the scale of the bottom figures is
divided by 20 with respect to the top ones.}}
\label{fig:traffic}

\end{figure*}

We consider the data used in Section 4 of \cite{Levy:Roueff:2009}, which corresponds to a recording
of 118 minutes of ADSL (Asymmetric Digital Subscriber Line) and \emph{Peer-to-Peer} (P2P) traffic
to which some TCP/SYN flooding type attacks have been added. As this data set does not contain full
routing information, it has been artificially distributed over a set of virtual monitors as
follows: the data is shared among $K=15$ monitors by assigning each source destination pair (source
IP address, destination IP address) to a randomly chosen monitor; a single monitor thus records all
the flows between two particular IP addresses. The experiments reported below are based on 50 independent replication of this process.
Finally, the existing anomalies have been down-sampled (by randomly dropping packets involved in the
attacks) to 12.5 and 25 packets/s, respectively, to explore more difficult detection scenarios.

Figure \ref{fig:traffic}-(a) displays the total number of TCP/SYN packets
received during each second by the different requested IP addresses. 
The number of TCP/SYN packets received by the four 
attacked destination IP addresses  are displayed in (d)
(12.5 packets/s case). 
As we can see from this figure, the first attack occurs at around 2000 seconds, the second at around 
4000 seconds, the third at around 6000 seconds and the last one at
around 6500 seconds. These attacks produce 33 ground-truth anomalies
-- abrupt increase or decrease of the signal. 
Figures \ref{fig:traffic}-(b), (c) display
the number of TCP/SYN packets globally exchanged within two
different monitors whereas (e), (f) 
focus on the traffic received by the attacked IP addresses within 
these two monitors.

The attacked IP addresses (bottom part of Figure~\ref{fig:traffic}) are completely hidden in the
global TCP/SYN traffic (top part of Figure \ref{fig:traffic}) and thus very difficult to
detect. Note also that 1006000 destination IP addresses are present in this data set, with an
average of 15000 destination IP addresses in each of the 118 one-minute observation windows. Hence,
real time processing of the data would not be possible, even at the monitor level, without a
dimension reduction step such as record filtering.

\subsection{Performance of the methods}

In what follows, the \emph{DTopRank} algorithm is used with the same
parameters as those adopted in \cite{Levy:Roueff:2009} for the
\emph{TopRank} algorithm, 
with one-minute windows divided in $P=60$ subintervals
of $\Delta=1$\;s, with $M=10$ and $S=60$. The parameter $d$ was set to $d=1$, due to the limited number of attacks expected in each one minute window.
In setting $P$ and $\Delta$ the main concern is the overall observation duration $\Delta \times P$
which should be sufficient to allow for meaningful statistical decisions while ensuring an
acceptable detection delay and that the extracted series can still be considered as stationary in
the absence of change. Note that the computational cost of the procedure also scales proportionally
to $P$. The influence of the other parameters ($d$, $M$ and $S$) is discussed at the end of Section
\ref{subsec:perf_synth}.

\begin{figure}[!ht]
\begin{center}
\includegraphics*[width=8cm, height=7cm]{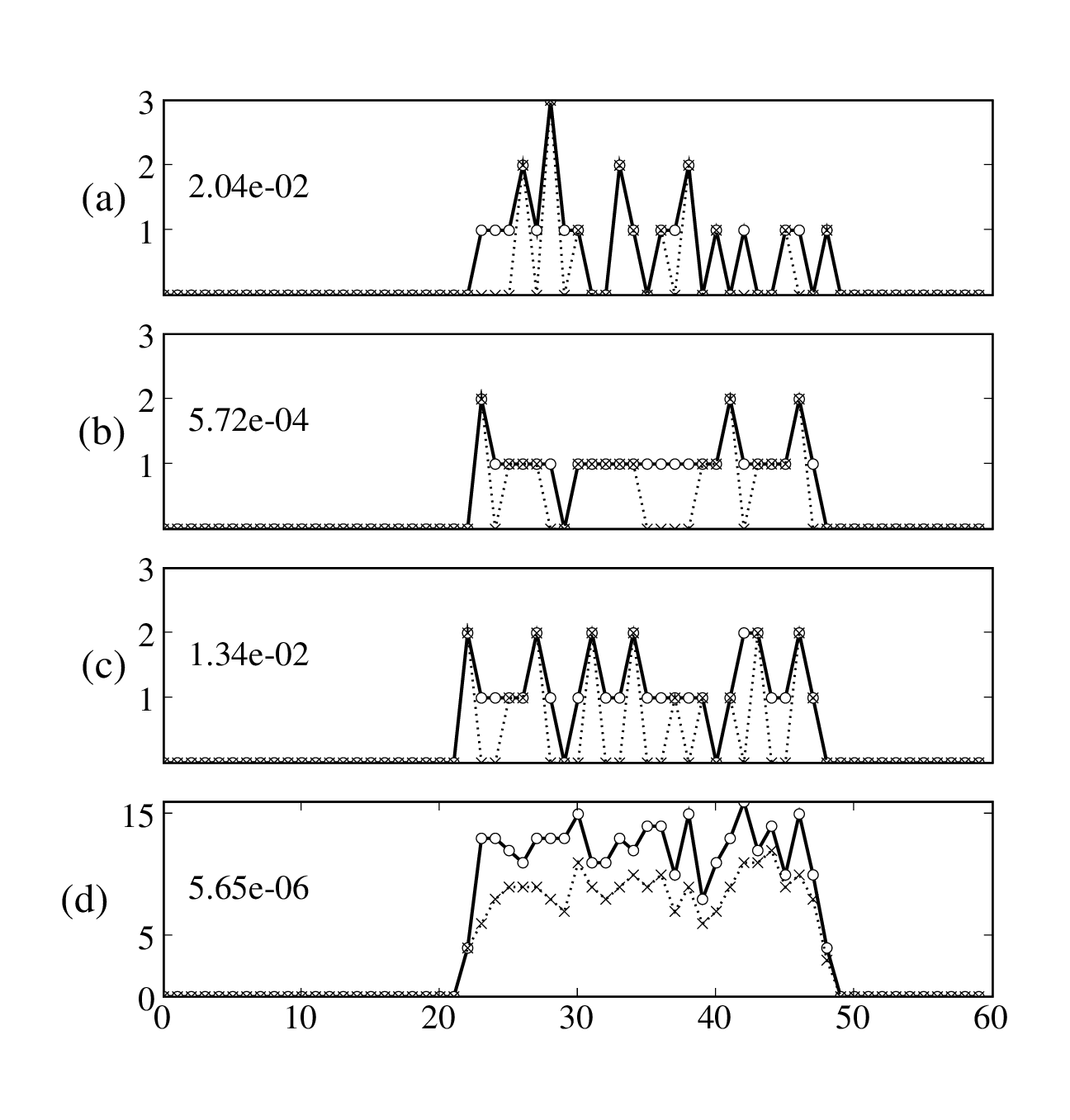}
\end{center}
\caption{\footnotesize{(a), (b), (c): times series
  $(\underline{X}_i^{(k)}(t),\;t=1,\dots,60)$ and
  $(\overline{X}_i^{(k)}(t),\;t=1,\dots,60)$ displayed with ('x') and
  ('o') respectively, for 3 different values of $k$, (d): $(\underline{Z}_i(t),\;t=1,\dots,60)$ and
$(\overline{Z}_i(t),\;t=1,\dots,60)$ displayed with ('x') and
  ('o') respectively. 
}
}
\label{fig:toprank}
\end{figure}

\begin{figure}[!ht]
\begin{center}
\hspace{-2mm}
\includegraphics*[width=.49\textwidth]{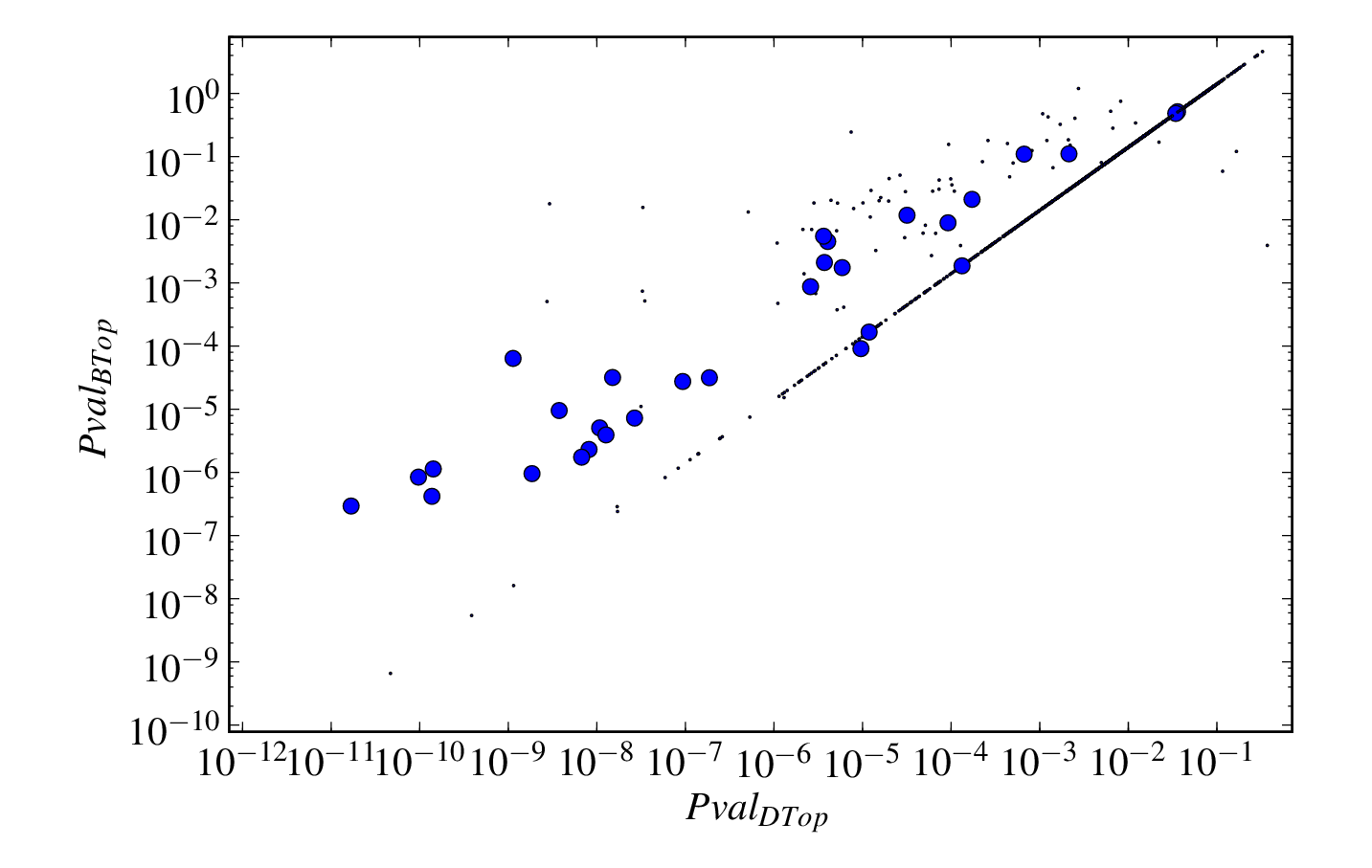}
\end{center}
\caption{\footnotesize{
    $(\textrm{Pval}_{\textrm{DTop}},\textrm{Pval}_{\textrm{BTop}})$
displayed with ('.') except for the ground-truth attacked IP addresses which
are displayed with ('$\bullet$').}}  
\label{fig:graph_color}
\end{figure}

Figure \ref{fig:toprank} and \ref{fig:graph_color} show the benefits of the
aggregation stage within the collector of the \emph{DTopRank}
algorithm with respect to the use of the simple Bonferroni correction in the 
\emph{BTopRank} algorithm.
Figures \ref{fig:toprank}-(a),(b) and (c) display the time series $(\underline{X}(t),\;t=1,\dots,P)$
and $(\overline{X}(t),\;t=1,\dots,P)$ associated to an attacked IP
address in three different monitors as well as the corresponding
$p$-values. Figure \ref{fig:toprank}-(d) displays the aggregated 
time series $(\underline{Z}(t),\;t=1,\dots,P)$ and
$(\overline{Z}(t),\;t=1,\dots,P)$, as defined in (\ref{eq:aggregation}),
as well as the associated $p$-value. Note
that the aggregated time series corresponds to the aggregation of 11
time series created by 11 different monitors where the attacked IP
address has been detected. The $p$-value of the aggregated time series is
much smaller than the ones determined at the local monitors, which enables
the detection of an attack which would be difficult to detect
within the local monitors.

Figure \ref{fig:graph_color} displays on the $x$ and
$y$-axes the quantities
$\textrm{Pval}_{\textrm{DTop}}$ and
$\textrm{Pval}_{\textrm{BTop}}$,
respectively. For a given IP address, $\textrm{Pval}_{\textrm{DTop}}$ corresponds to the $p$-value
computed with the \emph{DTopRank} 
and $\textrm{Pval}_{\textrm{BTop}}$ is obtained by applying the Bonferroni correction to the $p$-values transmitted by the monitors.
The \emph{DTopRank} provides smaller $p$-values than
the Bonferroni approach for IP addresses that were really attacked and
$p$-values of the same order as $\inf_{1\leq k\leq
  K}\textrm{Pval}_k$ for the other IP addresses.

\begin{figure}[!ht]
\begin{center}
\includegraphics*[width=.4\textwidth]{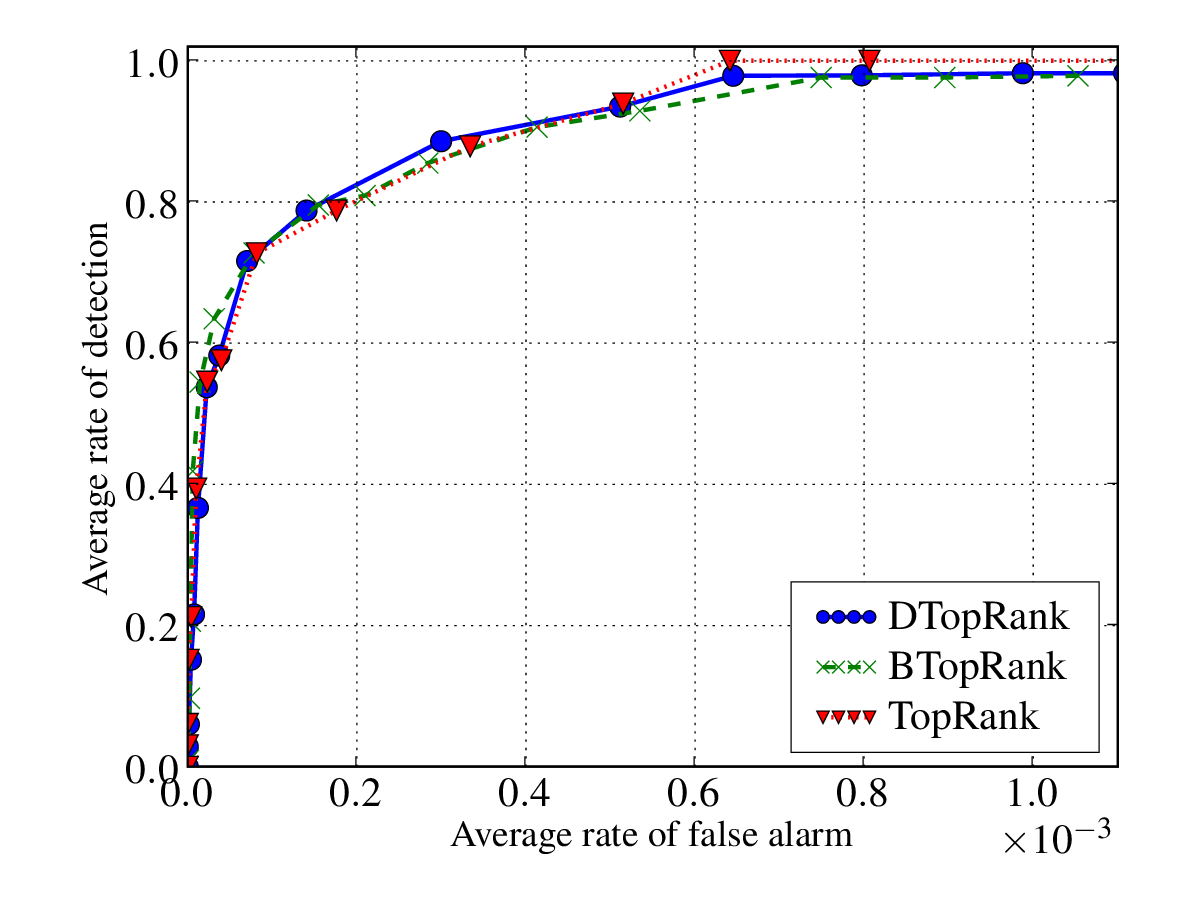}
\includegraphics*[width=.4\textwidth]{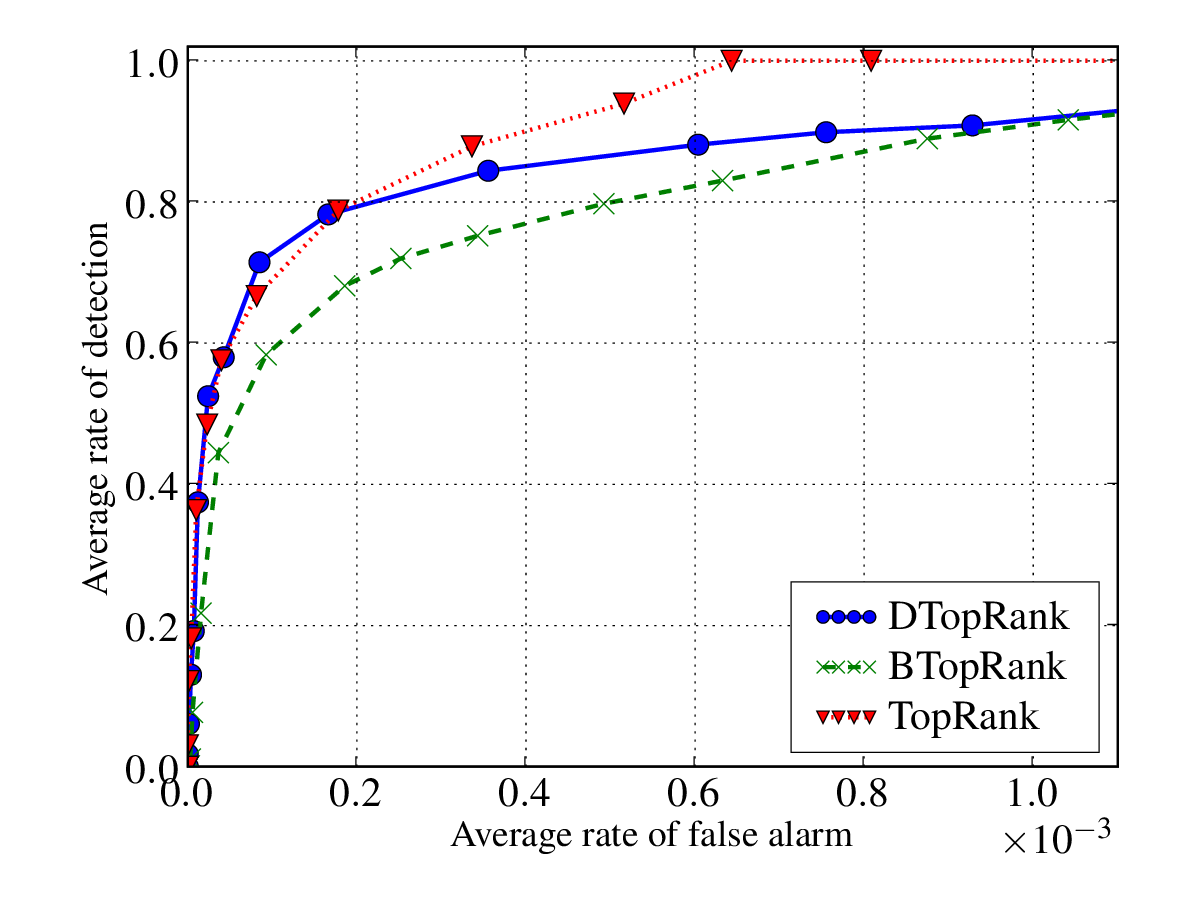}
\end{center}
\caption{\footnotesize{ROC curves for the
\emph{DTopRank}, \emph{BTopRank} and \emph{TopRank} algorithms
for attacks having intensities of 25 SYN/s (top)
and 12.5 SYN/s (bottom).}}  
\label{fig:top_bonf}
\end{figure}

The \emph{DTopRank} and \emph{BTopRank} algorithms are further compared
in Figure \ref{fig:top_bonf} which displays the ROC curves obtained
using these two methods with 50 Monte-Carlo replications in two
different cases. 
The bottom plot deals with attacks having
an intensity of 12.5 SYN/s. In the other situation, the attacks are the
same except that their intensity is 25 SYN/s. For comparison purpose, the
ROC curve associated to the non distributed \emph{TopRank} algorithm
is also displayed in both situations.
Figure \ref{fig:top_bonf} shows that for the 25 SYN/s-attacks, the three
methods give similar results. However, in the most difficult case of
the 12.5 SYN/s-attacks, the \emph{DTopRank} algorithm outperforms
the \emph{BTopRank} algorithm.

Thus, \emph{DTopRank} performs very similarly to the centralized algorithm, especially in the range
of interest where the false alarm rate is about 1e-4 (recall that there are about 15000 different
IP addresses in each one minute window). The quantity of data exchanged within the network is
however much reduced as the centralized algorithm needs to obtain information about, on average,
34000 flows per minute whereas the \emph{DTopRank} algorithm only need to transmit the $d$ upper
and lower censored time series from the monitors to the collector. For $d=1$ and $K=15$, this
amounts to 1800 scalars that need to be transmitted the collector, versus $34000 \times 5$ (start
and end time stamps, source and destination IP, number of SYN packets for each flow) for the
centralized algorithm, resulting in a reduction of almost two orders of magnitude of the data that
needs to be transmitted over the network.


\section{Application to synthetic  data}\label{sec:simul}

In this section, we provide results obtained on simulated data with two specific goals in
mind. First, the the traffic trace used in Section \ref{sec:real} contains generated attacks but is
not fully labeled. Hence, it could be the case that non-labeled anomalies are already present in
the background ADSL and P2P traffic contributing to a slight overestimation of false alarms (see
\citealp{Levy:Roueff:2009}). Second, the random decentralization approach used in Section
\ref{sec:real} does not necessarily correspond to a realistic network topology. In this section, we
thus consider synthetic high-dimensional data corresponding to an idealized minute of traffic
containing a single anomaly, as measured by 15 monitors randomly positioned on a plausible network
topology.

\subsection{Description of the data}\label{sec:descr:synth}

\begin{figure}[!ht]
  \includegraphics[scale=0.3]{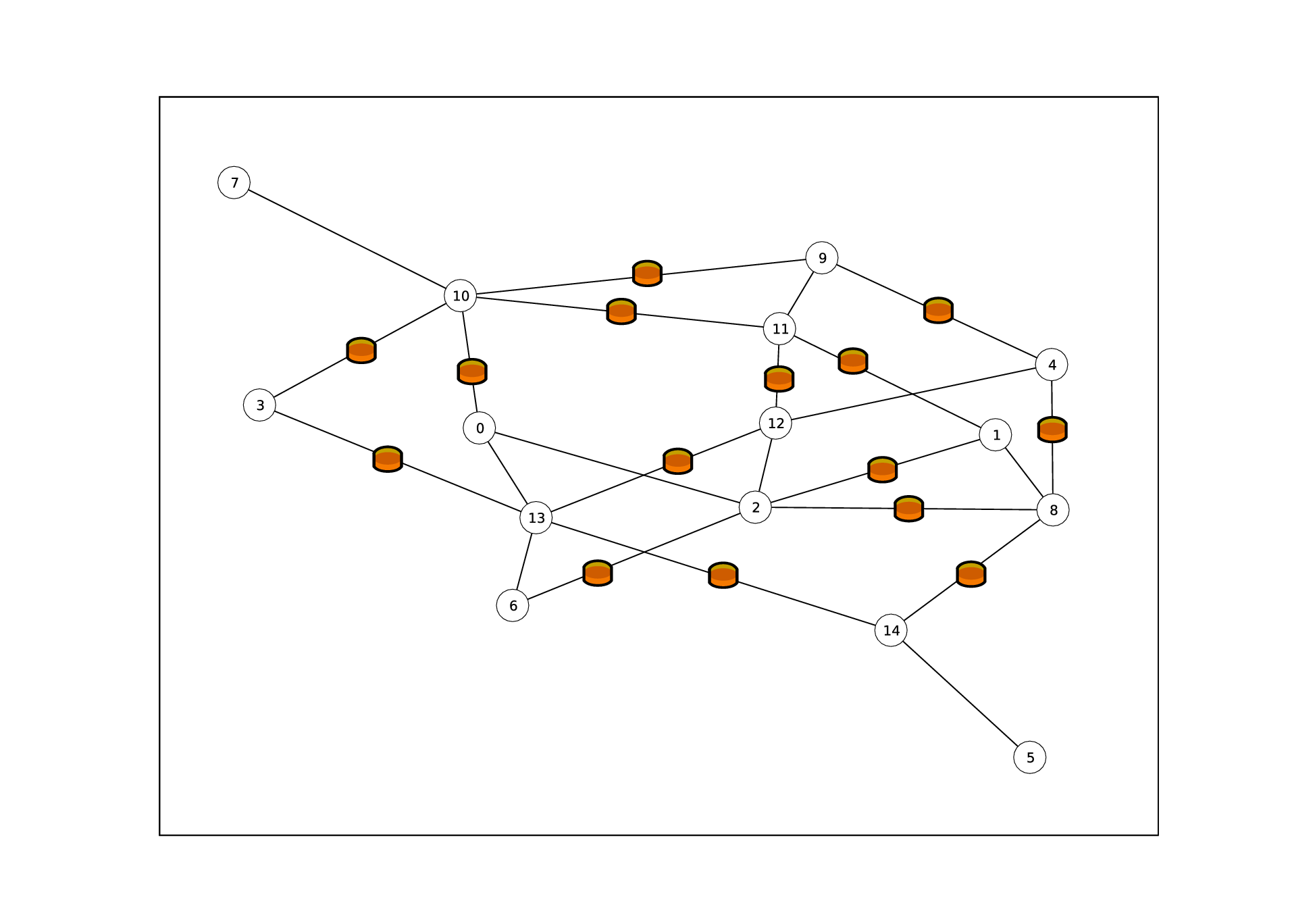}
  \centering
  \caption{Generated graph: nodes are displayed with circled numbers
    and monitors with colored boxes.}
\label{fig:graph}
\end{figure}

A network topology is generated in which synthesized traffic between
hosts located in the nodes of that network is injected.
We first generate an Erd\H{o}s-Rényi 
random graph (\cite{erdos:1959})
with 15 nodes
and a probability of edge creation of $0.15$.
The generated graph is displayed in Figure \ref{fig:graph}. It is similar in terms of number of nodes or nodes degrees
to the Abilene network, which has been widely considered in the context of network
anomaly detection, see \cite{Lakhina:2004} and
\cite{huang:nips}.
This graph has been generated once and is used for all replications
of the Monte-Carlo simulations that will follow.
For each Monte-Carlo replication, a node of the graph is randomly assigned to each
of the $D=1000$ IP addresses and $K=15$ monitors are also randomly
positioned on 15 of the 24 edges of the graph, see Figure \ref{fig:graph}.

Using the shortest path \textsf{\cite{dijkstra:1959}} algorithm, the routes
between each node of the network are computed, that is the lists of
edges of the graph that form the path between the nodes.
These routes are used to determine which monitors will see the traffic 
between two hosts. Note that in our procedure, we have deliberately
not considered network links capacity, that would otherwise imply some 
more sophisticated dynamic routing algorithms, which is beyond the scope of this contribution.

The traffic injected in this network is generated as follows.
For a given Source-Destination IP address pair $(i,j)$, we follow \cite{Levy:Roueff:2009} and model the SYN packet traffic
using a Poisson point process with a given intensity $\theta_{i,j}$, expressed as the number of SYN packets
received by $j$ per sub-interval of the observation window.
In Network applications, different Source-Destinations pairs exchange a very different amount of traffic.
Hence we shall use different intensities for each pair of hosts.
To take into account this diversity, we propose using the realizations of a Pareto distribution
for the parameters of the different intensities so that a lot of
machines receive a small number of SYN packets while a few receive a lot.
Note that \cite{nucci:sridharan:taft:2005:sigcomm} similarly use a heavy-tailed
distribution to generate network traffic.

We first randomly generate a sequence $(\mu_{k})_{1\leq k\leq N}$ of
intensities with the Pareto distribution having the following density: 
$\gamma \alpha /(1+\gamma x)^{1+\alpha}$, when $x>0$, with $\alpha =
2.5$ and $\gamma = 0.72$, which roughly corresponds to what we
observed in the (centralized) real traffic traces used in Section \ref{sec:simul}.
The parameters $\mu_{k}$ are assumed to be sorted as follows:
$\mu_1\geq\dotsb\geq\mu_N$. 

Here, $(X_{i,j}(t))_{1\leq t\leq P}$ correspond to the number of SYN
packets sent by $i$ and received by $j$ in each of the $P$
sub-intervals of the observation window, where $i,j$ are in
$\{1,\dotsc,D\}$. Among these $N$ time series, $N_a$ of them 
correspond to the traffic received by the attacked destination
IP address $j_0$, which is assigned to a fixed location, in node 7, at the ``edge''
of the network (see Figure \ref{fig:graph}).
This traffic, which is sent by source
IP addresses $i$ belonging to a randomly chosen subset $\mathcal{I}_a$
of $\{1,\dotsc,D\}$, is generated as follows:
\begin{equation*}
  \label{eq:paremetersSimulationAnomalies}
  \forall i \in \mathcal{I}_{a}, (X_{i,j_{0}}(t))_{1\leq t\leq \tau} 
  \iid \textrm{Poisson}(\theta_{i,j_{0}}) \; ,
\end{equation*}
and
 \begin{equation*}
  \forall i \in \mathcal{I}_{a}, (X_{i,j_{0}}(t))_{\tau < t\leq P} \iid \textrm{Poisson}(\eta
  \theta_{i,j_{0}})\; ,
\end{equation*}
where $\eta$ is a positive number which modulates the change intensity, $\tau$ is the change-point
instant and $(\theta_{i,j_0})_{i\in\mathcal{I}_{a}}$ are chosen in $(\mu_{k})_{40 N_a\leq k\leq 41
  N_a}$. $(\theta_{i,j_0})_{i\in\mathcal{I}_{a}}$ are thus chosen around 0.6 (0.4-quantile of the
Pareto distribution with parameters $\alpha$ and $\gamma$, whose mean is about 0.93). Hence, the
attack to be detected consists of a multiplicative increase in intensity of $N_a$ attacker sources,
whose intensity is otherwise in the bulk of the distribution of the intensity (close to the
0.4-quantile). The remaining background traffic is generated as:
\begin{equation*}
  \label{eq:parametersSimulation}
  \forall i \in \{1,\dots,D\},\; j\neq j_{0},\;  (X_{i,j}(t))_{1\leq
    t\leq P} \iid \textrm{Poisson}(\theta_{i,j})\; ,
\end{equation*}
where $(\theta_{i,j})_{i\in \{ 1,\dotsc,D \}, j\neq j_{0}}$ are chosen
randomly in the remaining values of $\mu_{k}$: $(\mu_{k})_{k\notin[40 N_a;41 N_a]}$.



\begin{figure*}[!ht]
  \centering
  \begin{tabular}{ccc}
    \hspace{-8mm}\includegraphics[width=.3\textwidth]{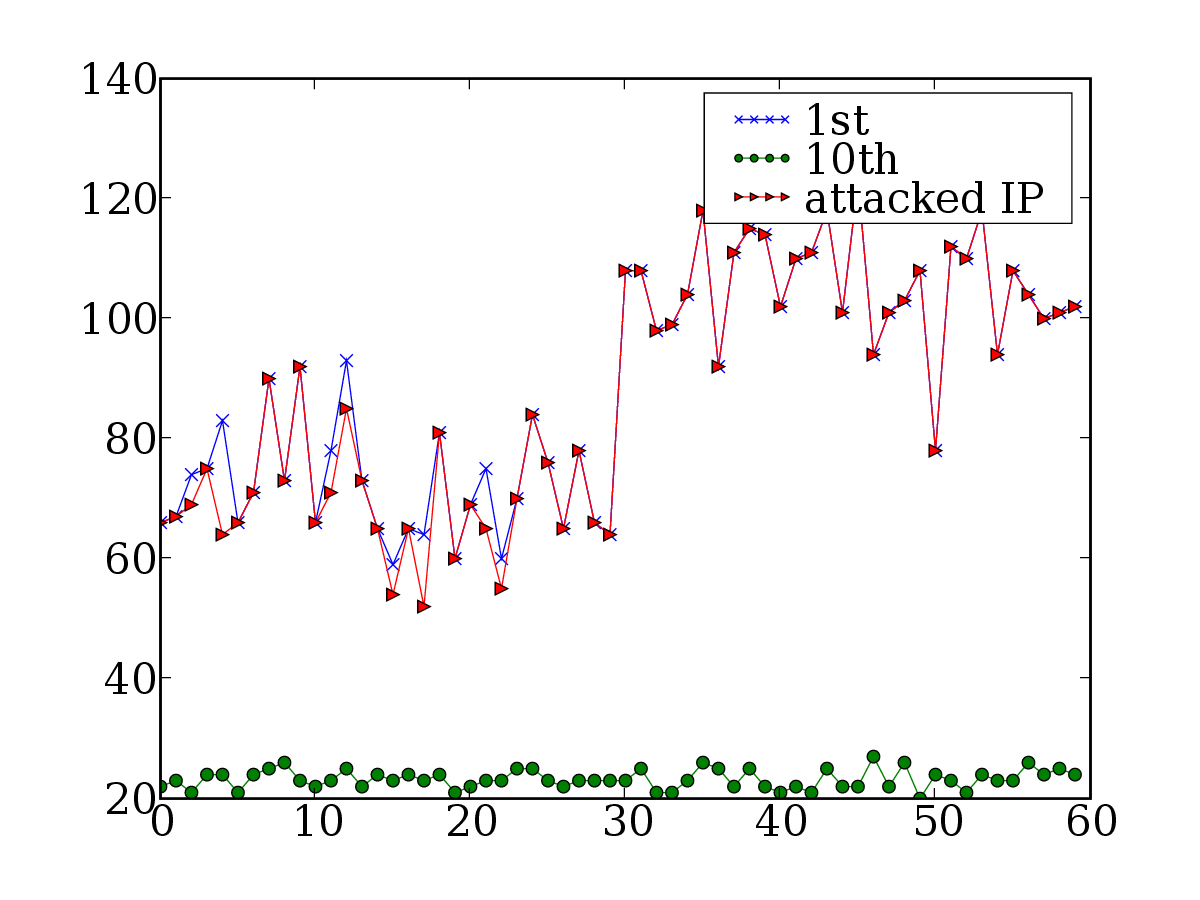} &
    \hspace{-6mm}\includegraphics[width=.3\textwidth]{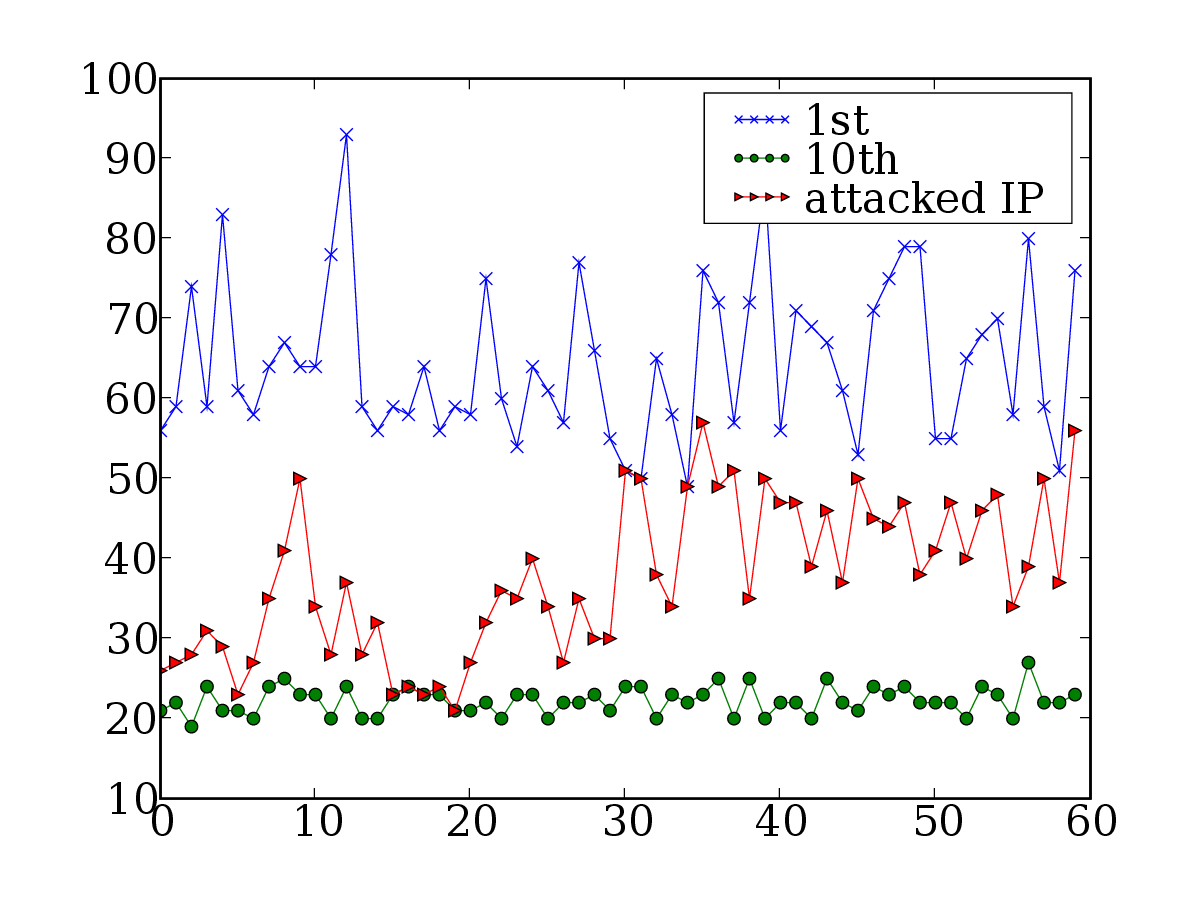} &
    \hspace{-6mm}\includegraphics[width=.3\textwidth]{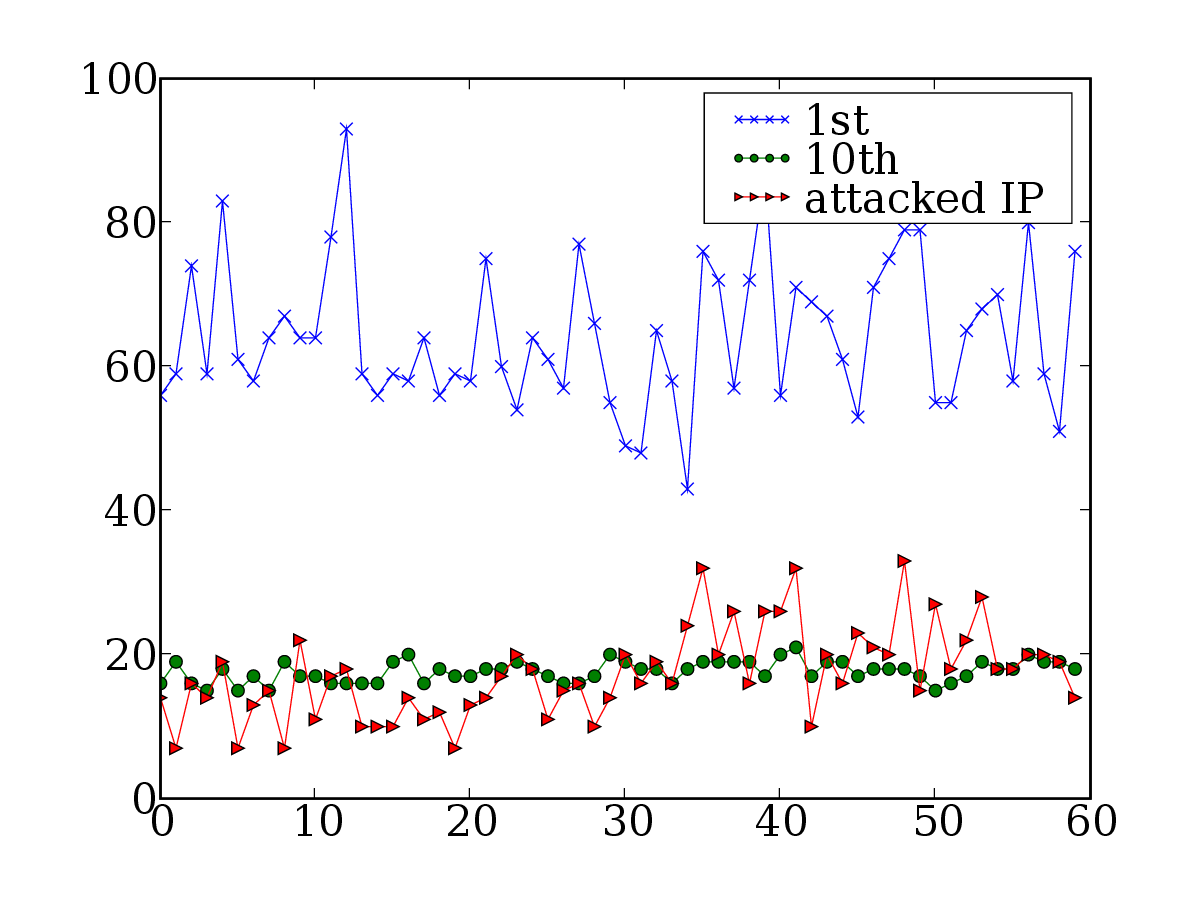} \\
    
   \hspace{-8mm} (a) &\hspace{-6mm} (b) &\hspace{-6mm} (c) \\

    \hspace{-8mm}\includegraphics[width=.3\textwidth]{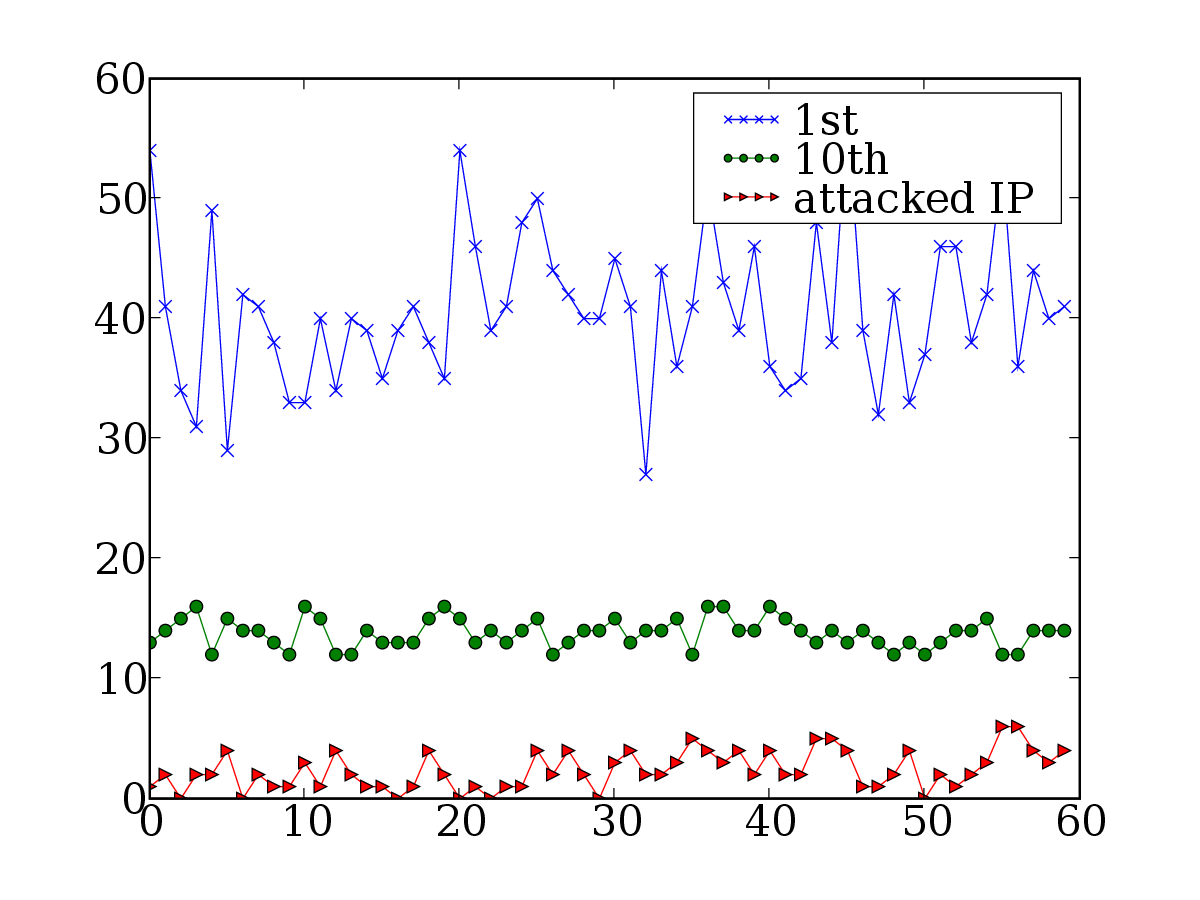} &
    \hspace{-6mm}\includegraphics[width=.3\textwidth]{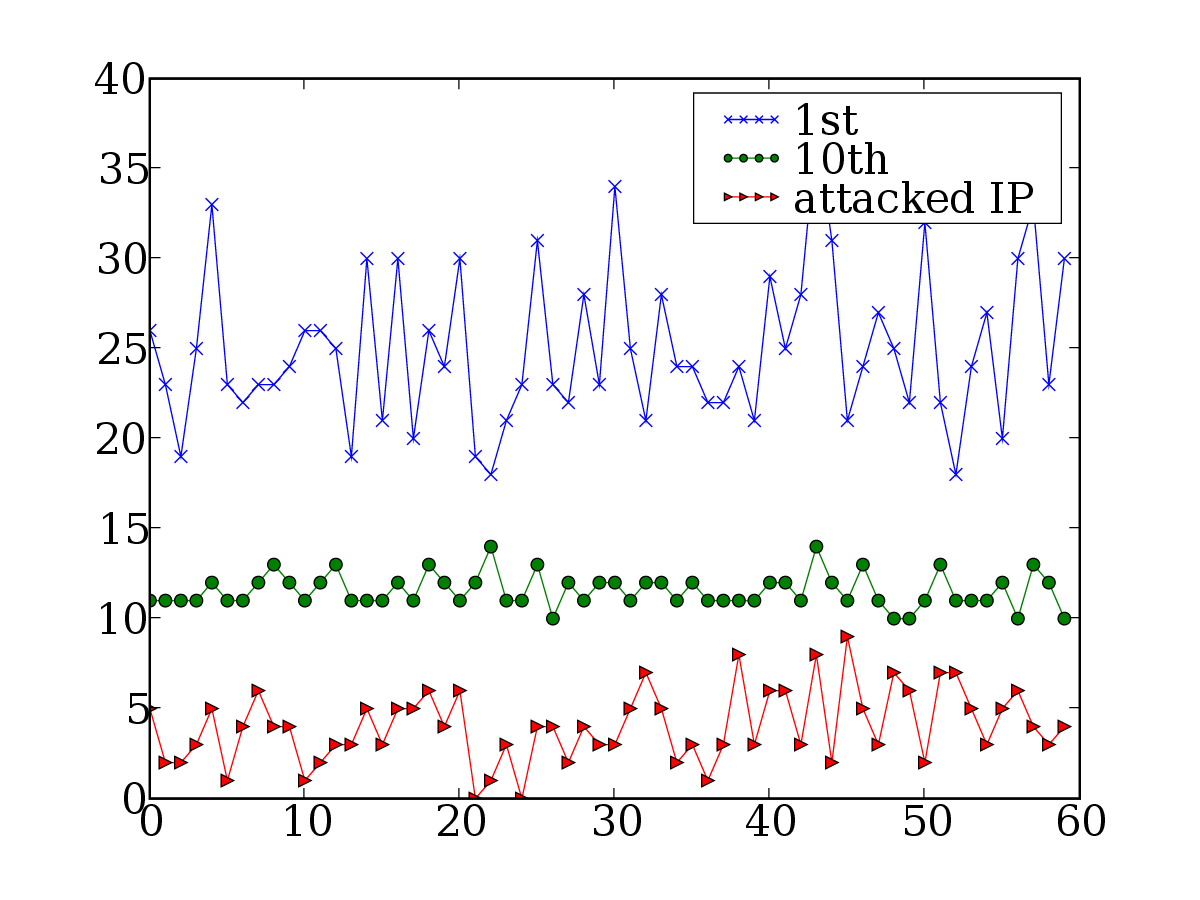} &
    \hspace{-6mm}\includegraphics[width=.3\textwidth]{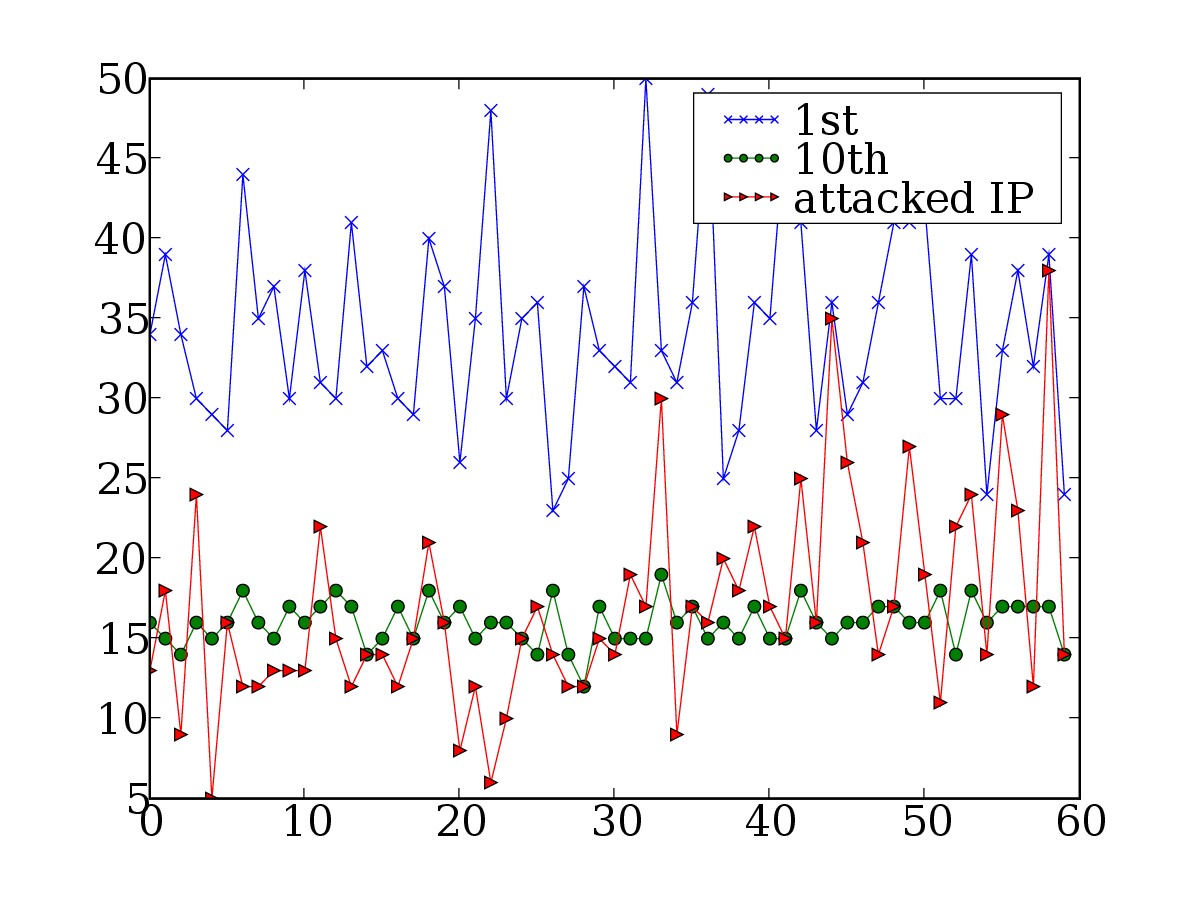} \\

 \hspace{-8mm}   (d) &\hspace{-6mm}(e)&\hspace{-6mm}(f) \\

    \hspace{-8mm}\includegraphics[width=.3\textwidth]{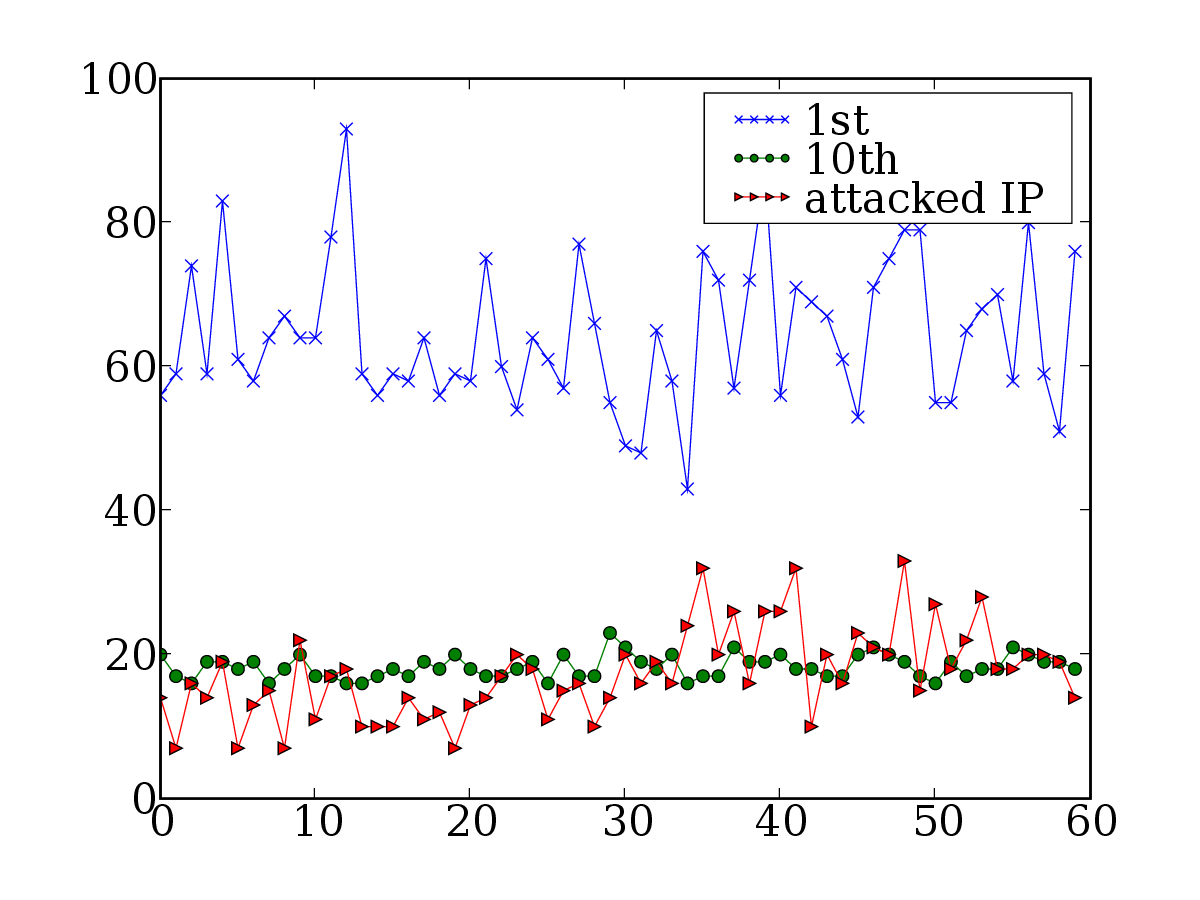} &
    \hspace{-6mm}\includegraphics[width=.3\textwidth]{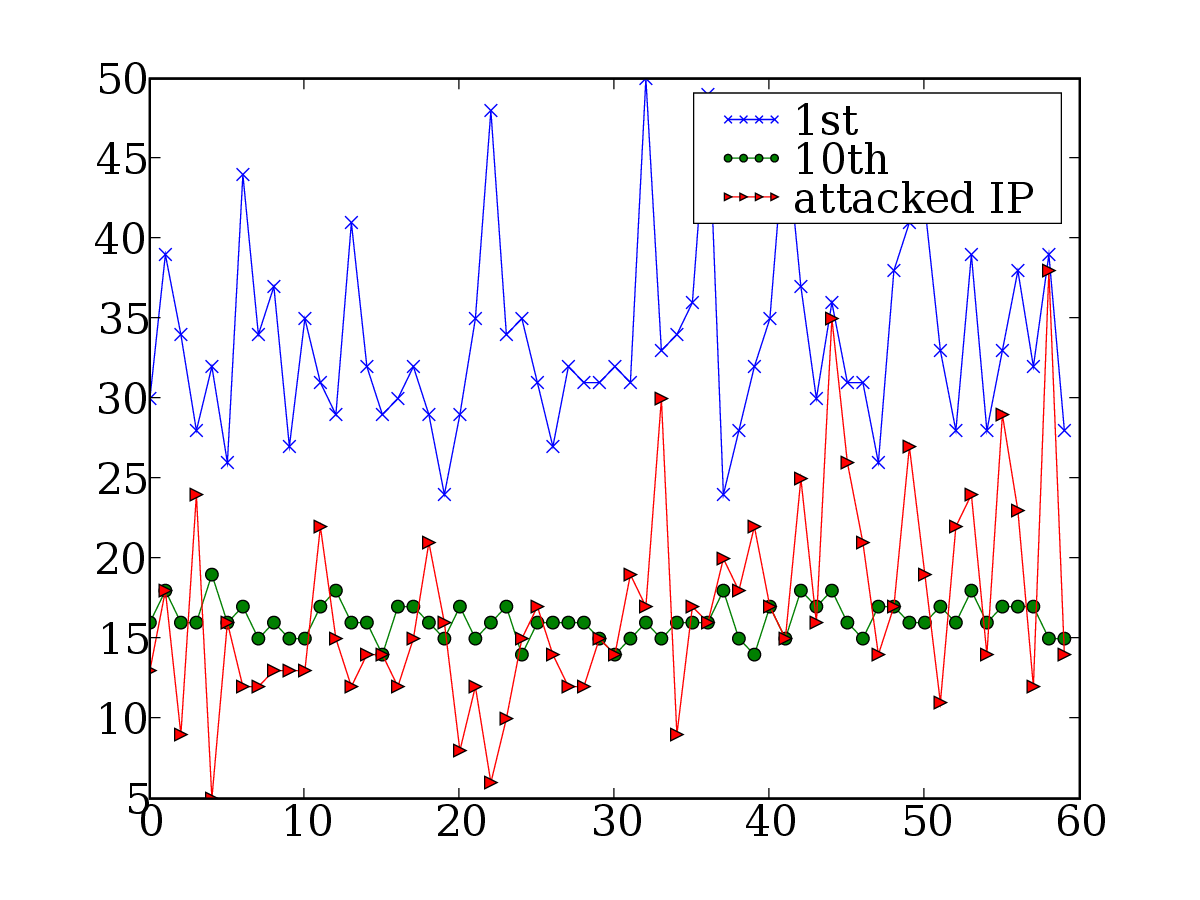} &
    \hspace{-6mm}\includegraphics[width=.3\textwidth]{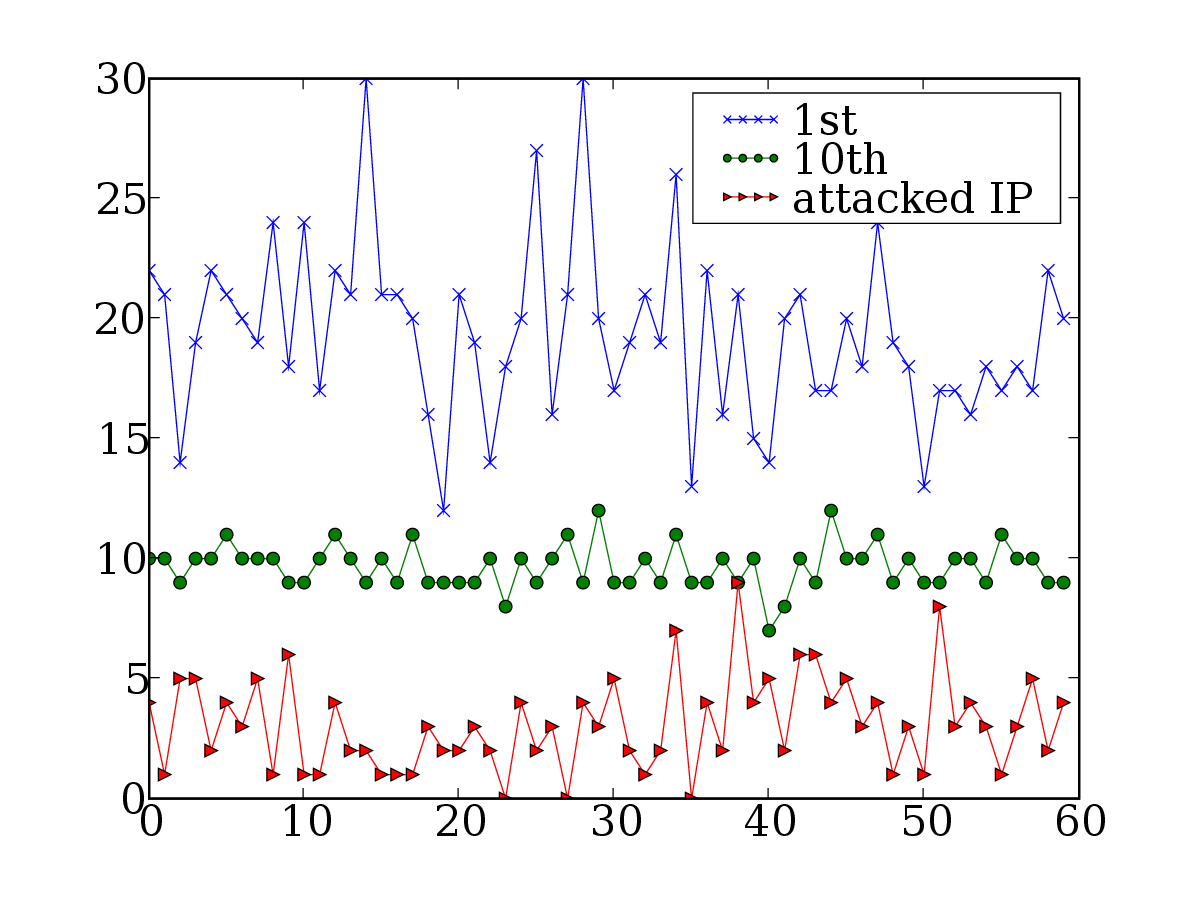} \\

  \hspace{-8mm}  (g)&\hspace{-6mm}(h)&\hspace{-6mm}(i)
  \end{tabular}
  \caption{Time series formed in 9 monitors by the number of packets 
received by the first (``$\times$'') and 10th (``$\bullet$'') most solicited destination IP address
at each sub-interval, as well as the time series of the attacked address $j_{0}$ (``$\triangleright$'').}
  \label{fig:sim_ts}
\end{figure*}

In the experiments below, $N=10100$, $N_{a}=100$, $P=60$, $\tau=30$
and we consider different values for the parameter $\eta$ ($1.2, 1.5$) in order
to modulate the detection difficulty.
Results presented in Figures~\ref{fig:sim_ts} to~\ref{fig:roc_sim} correspond to the case where $d=1$ and the influence of
  $d$ on the performance is discussed in the final paragraph of Section
  \ref{subsec:perf_synth}.
With these settings, the  DDoS-type attacks against $j_{0}$ are 
generated by a large number $N_{a}$ of source hosts
coming from all routing nodes in in the network. Hence, these attacks can be locally
(within a monitor) very difficult to distinguish from the background
traffic, as can be seen in Figure \ref{fig:sim_ts}:
It displays for each monitor, when $\eta=1.5$, an example of the time series formed by the number of packets 
received by the first (``$\times$'') and 10th (``$\bullet$'') most solicited destination IP address
at each sub-interval as well as the time series of
the attacked address $j_{0}$ (``$\triangleright$'').
The monitors that have not detected any traffic directed to $j_{0}$ were omitted.
In (d), (e) and (i), $j_{0}$ is detected by the monitor, but the
number of packets is never high enough to be
selected by the record filtering step and to appear in
$\{\mathcal{T}_M(t),\; t=1,\dotsc,60\}$.
Hence in these monitors, no change detection test is performed for $j_{0}$.
In the other six figures, all steps of the \emph{TopRank} algorithm
are carried out since the number of packets sent to $j_{0}$ is high enough.
A special case is shown in (a), which displays the time series in the monitor located on the edge between
nodes 10 and 7, see Figure \ref{fig:graph}, which is the link where all the traffic directed to the attacked IP address $j_{0}$ appears.

\subsection{Performance of the methods}\label{subsec:perf_synth}

\begin{figure*}[!ht]
  \begin{center}
    \begin{tabular}{cc}
      \includegraphics[width=.4\textwidth]{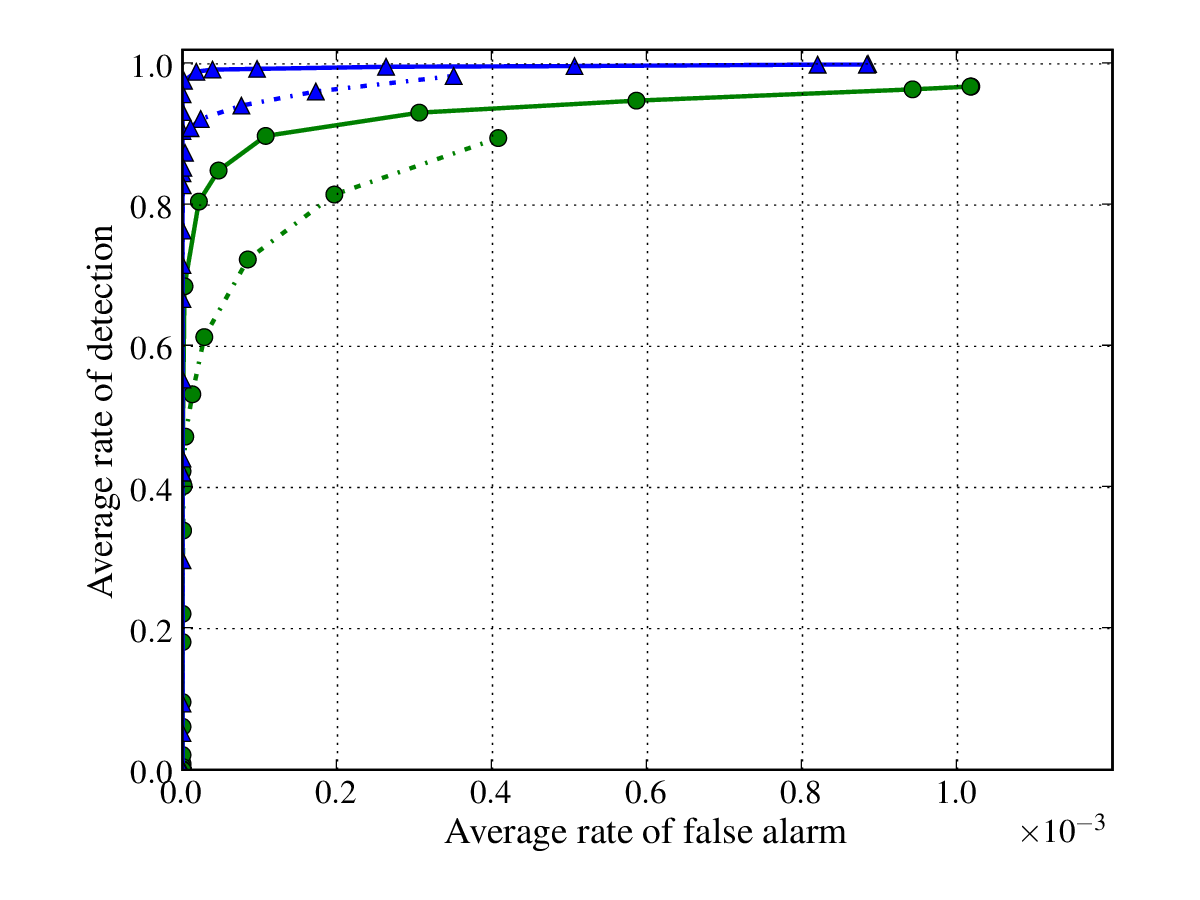}&
      \includegraphics[width=.4\textwidth]{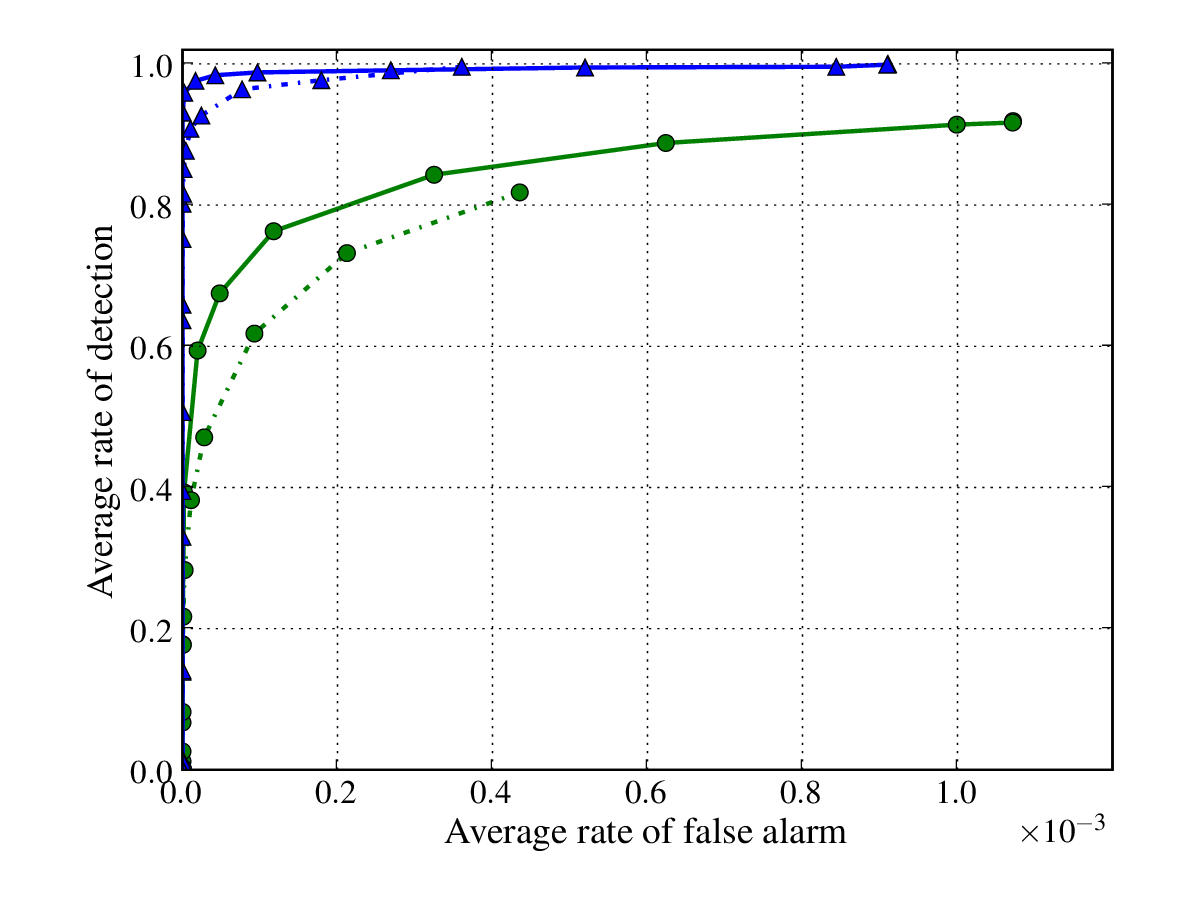}
    \end{tabular}
    \caption{Left: ROC curves for \emph{DTopRank} (solid lines) and \emph{BTopRank} (dashed lines),
      for $\eta=1.2$ (``$\bullet$''), $1.5$ (``$\blacktriangle$''). 
      Right: similar simulation when forbidding the 10-7 edge from the monitors.
    }
    \label{fig:roc_sim}
    
  \end{center}
\end{figure*}

The two methods described in Section \ref{sec:description} are compared by 
computing their false alarm and detection rates when tested on 1000 Monte-Carlo 
replications of the synthetic data described in Section
\ref{sec:descr:synth}.
The left plot of Figure \ref{fig:roc_sim} displays the corresponding ROC curves for
different values of $\eta$ (1.2 and 1.5); the solid and dashed lines show
the results of the \emph{DTopRank}  and \emph{BTopRank} algorithms, respectively.

For larger values of $\eta$ (1.5), both methods perform very well,
with a few missed attacks for a low false alarm rate. \emph{DTopRank} yields slightly better results than
the other method.
For $\eta=1.2$ for which attacks are more difficult to detect, the detection
performance is naturally lower for both algorithms; the toll is however heavier on
\emph{BTopRank} than on \emph{DTopRank}.

We observed that the detection performance was improved for Monte Carlo runs in which a monitor is
assigned to the 7--10 edge of Figure~\ref{fig:graph}. Indeed in this case, at least a monitor has
access to all the traffic sent to the attacked IP address sitting at node $j_{0}=7$. The right plot
of Figure \ref{fig:roc_sim} corresponds to the case where this configuration is avoided in the
Monte Carlo simulation, which gives some idea of the significance of the phenomenon. For a given monitor
topology, the detection performance is thus better for target addresses located at the edge of the
network, behind a monitor. In the opposite case however, the detection performance is still
appreciable due to the aggregation, at the collector level, of the information sent by the
monitors.

\begin{figure*}[!ht]
  \begin{center}
    \begin{tabular}{ccc}
      \hspace{-10mm}
      \includegraphics[width=.38\textwidth]{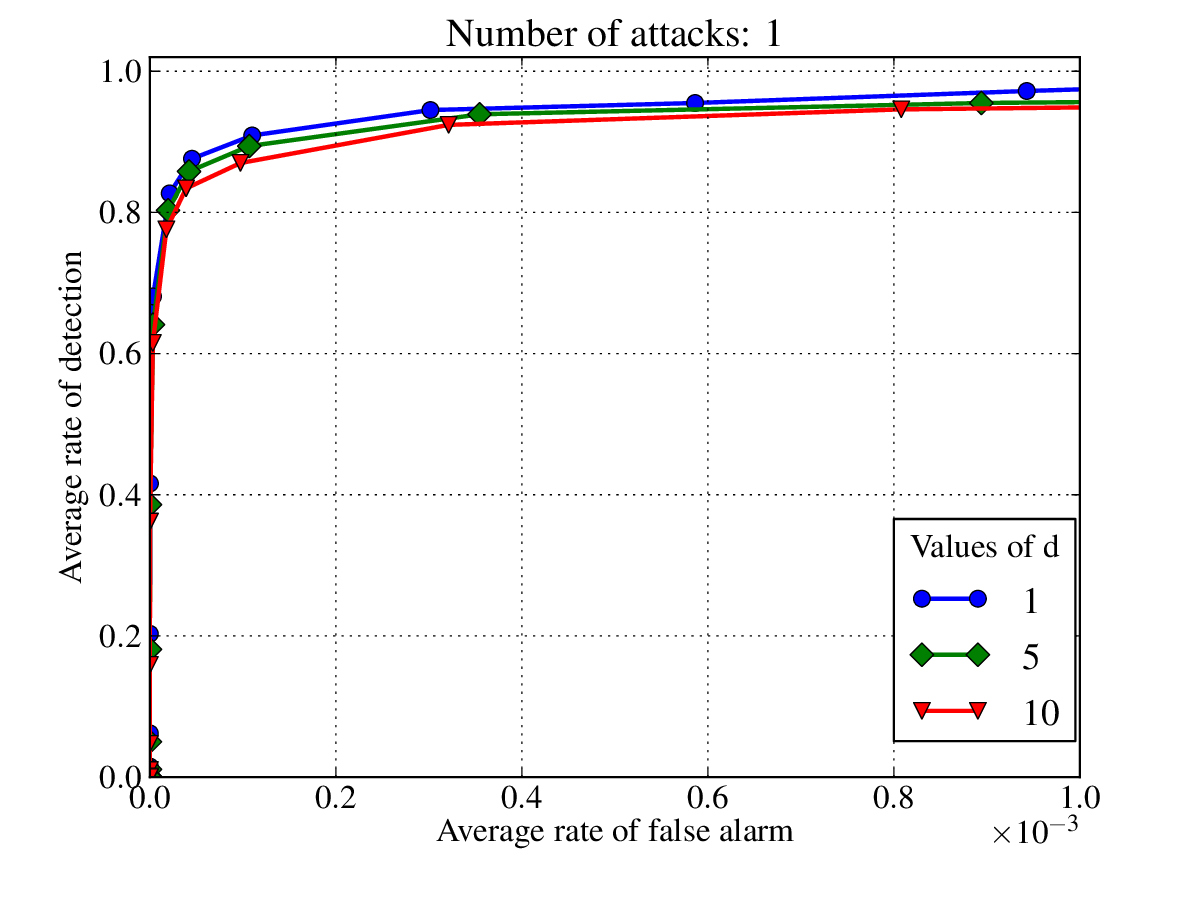}&
      \hspace{-10mm}
      \includegraphics[width=.38\textwidth]{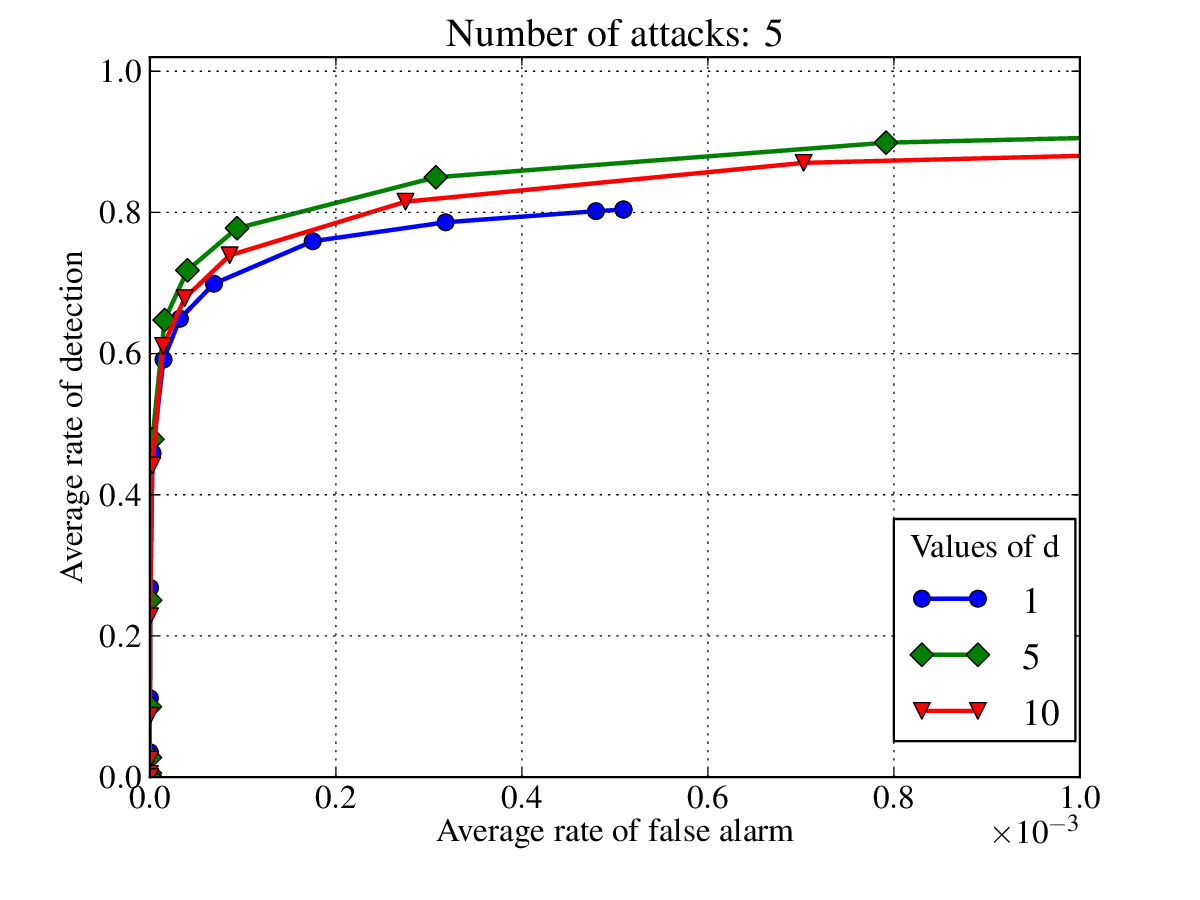}&
      \hspace{-10mm}
      \includegraphics[width=.38\textwidth]{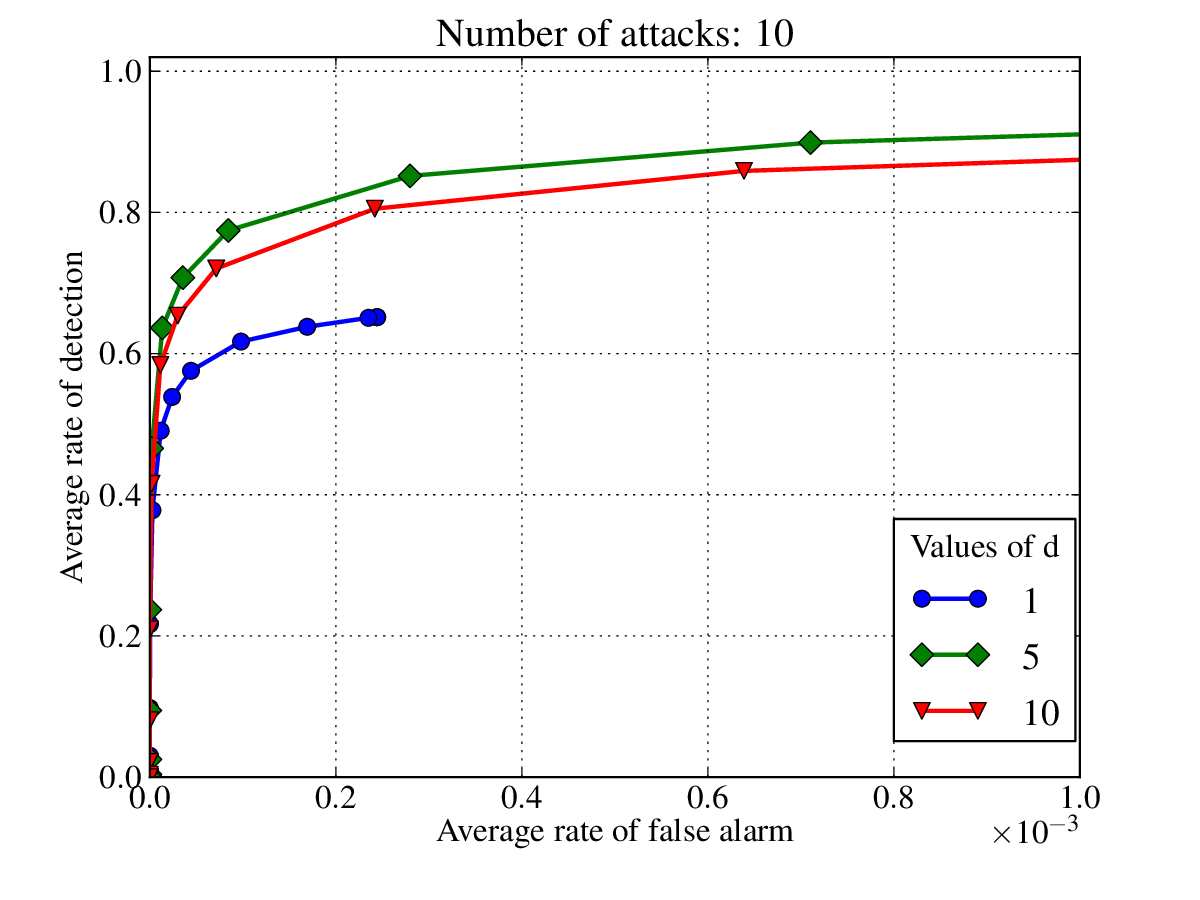}
    \end{tabular}
    \caption{ROC curves for the \emph{DTopRank} algorithm when the
      number of attacks per observation window is 1, 5, 10 (from left
      to right) and when $d$ equals to 1 (``$\bullet$''), 5
      (``$\blacklozenge$''), 10 (``$\blacktriangledown$'').}
    \label{fig:roc_sim_multiattack}
   
  \end{center}
\end{figure*}

We now consider the influence of the parameter $d$ on the performance of
\emph{DTopRank}. Figure~\ref{fig:roc_sim_multiattack} displays the ROC curves associated to the
\emph{DTopRank} algorithm for different values of $d$ ($d$=1, 5, 10) and a varying number of
attacks per observation window.  From this figure, one can first observe from the leftmost plot
that using $d=1$ for the \emph{DTopRank} algorithm is optimal when there is at most one attack per
window but is comparatively less advantageous as the number of attacks per window
increases. Indeed, the collector then receives the time series associated to, at most, one IP
address per monitor, yielding low detection rates. Increasing the value of $d$ thus improves the
performance of the \emph{DTopRank} algorithm when several attacks occur in the same window. However,
the rightmost plot of Figure~\ref{fig:roc_sim_multiattack} reveals a form of plateauing and $d=5$
appears to provide the best tradeoff, even in cases where the actual number of attacks to be
detected per window is larger than 5. This behavior can be explained by the observation that the
traffic directed towards a particular attacked IP address is not always visible by all local
monitors.

\begin{figure}[h]
  \centering
  \includegraphics[width=.4\textwidth]{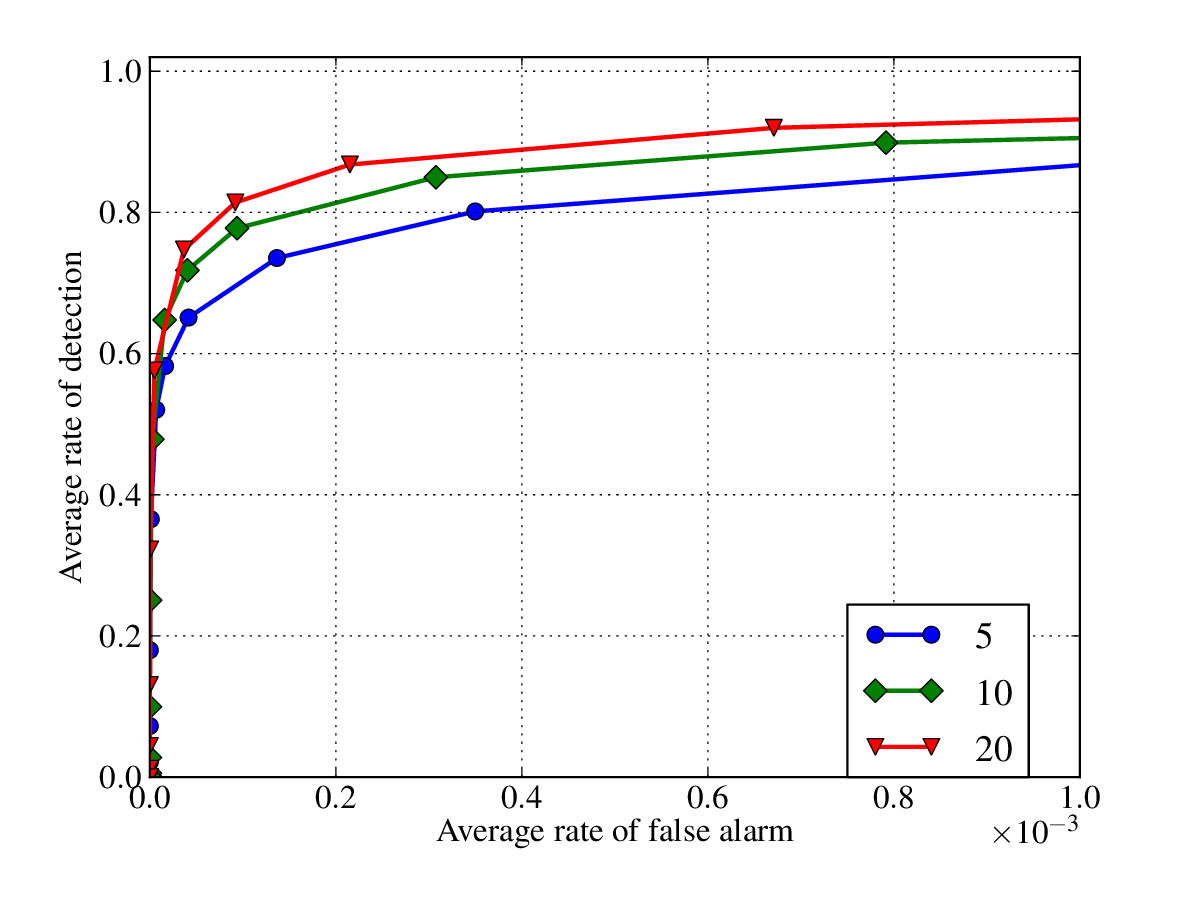}
  \caption{ROC curves for the \emph{DTopRank} algorithm when the
    parameter $M$ takes the values of 5 (``$\bullet$''), 10
    (``$\blacklozenge$'')  or 20 (``$\blacktriangledown$'').}
  \label{fig:roc_influence_M}
\end{figure}

We have also investigated how the parameters $M$ 
and $S$ (see Section
\ref{subsec:loc_proc})
affect the performance of the \emph{DTopRank} algorithm.
Figure \ref{fig:roc_influence_M} shows the impact of $M$ while keeping
$S$ constant to the value of $60$; both the number of attacks per
observation window and the parameter $d$ were set to 5. Similar experiments,
whose results are not reported here, have also shown that changing $S$ to values of 30 or 120 does not
affect at all the ROC curves.
From Figure \ref{fig:roc_influence_M}, one can observe that larger values of $M$ consistently
yield improved results, which is related to a lower 
censorship. Recall however that both the computational cost and the memory footprint within each monitor are proportional to $M$. Hence, the choice of $M$ results from a tradeoff between performance and memory or CPU-consumption.
The absence of influence of $S$ may be explained by the fact that only $d$ values are sent to the
central collector by the monitor. It suffices that $S$ be much larger than $d$ and of the order of
$P$ so as to record all the ``heavy hitters'' that correspond to the largest $N_i(t)$, for all
$t\in\{1,\dotsc,P\}$.

\section{Conclusion}

In this paper, we proposed a distributed method for detecting and localizing DDoS attacks in 
Internet traffic. With this approach, a local processing based on a record filtering technique
followed by a nonparametric rank test is performed within the local monitors. Only the censored
time series of IP addresses corresponding to the smallest $p$-values are transmitted and aggregated
in the collector.  The central decision, at the
collector level, is based on a non-parametric change-point test for two way censored data. 
The processing carried out both in the monitors and in the central collector is
sufficiently simple to make real-time implementations possible.
The proposed algorithm has been shown to reveal
attacks which are not locally detectable. Indeed, the statistical performance of the 
\emph{DTopRank} algorithm is close to that achieved by the fully centralized detector but with
a greatly reduced communication overhead.
An additional interesting feature of the proposed aggregation and detection mechanism is the fact
that it operates similarly at the monitor and collector level. Hence, the test could also be applied
hierarchically, with tree structured monitors, so as to produce decisions corresponding to groups
of monitors of different granularity in the network.

\appendix

\section{Appendix}


\subsection{Proof of Theorem \ref{theo:null_hyp}}
\label{sec:propH0}







The following proof is based on \cite[Theorem 24.2]{billingsley:1968},
which asserts that if $\xi_1,\dotsc,\xi_n$ are exchangeable random
variables (each permutation of the set of variables has the same joint
distribution) and satisfy, as $n\rightarrow \infty$,
\begin{equation}\label{eq:cond_billingsley}
\sum_{i=1}^{n}\xi_{i} \inproba 0,\quad
\sum_{i=1}^{n}\xi^2_i \inproba 1,\quad
\max_{1\leq i\leq n} |\xi_{i}| \inproba 0\;,
\end{equation}
then  $\{\sum_{i=1}^{\lfloor nt\rfloor}\xi_{i}\; ,  0 \leq t \leq 1\}
 \inlaw \{B(t)\;,\; 0 \leq t \leq 1\}$, as $n\rightarrow \infty$,
where $B$ is a Brownian bridge.

We apply this theorem to the random variables
$Y_1,\dotsc,Y_P$, defined in (\ref{eq:Y_i}), which are exchangeable since
$(\underline{X}(i),\overline{X}(i))_{1\leq i\leq P}$ are i.i.d random vectors.
Let us now check the three conditions in (\ref{eq:cond_billingsley}).
By the anti-symmetry of the kernel $h$,
$$
\sum_{i=1}^{P}U_{i}=\sum_{i=1}^{P}\sum_{j=1}^{P}h(i,j)
    =  0 \;,
$$
which gives the first condition of the theorem.
The second one follows from the definition of $Y_{i}$:
$$
   \sum_{i=1}^{p}Y_{i}^{2}  =  \inv{\sum_{j=1}^{p}U^{2}_{j}} \sum_{i=1}^{p}U_{i}^{2} = 1 \;.
$$
To check the third condition, denote by
$F_{P}$ (resp. $G_{P}$) the empirical c.d.f. of
$\overline{X}(1),\dotsc,\overline{X}(P)$
(resp. $\underline{X}(1),\dotsc,\underline{X}(P)$):
\begin{equation}
  \label{eq:ecdf}
  F_{P}(t) = P^{-1}\sum_{i=1}^{P}\1(\overline{X}(i)\leq t) \textrm{
    and }   G_{P}(t) = P^{-1}\sum_{i=1}^{P}\1(\underline{X}(i)\leq  t)\; .
\end{equation}
Note that
\begin{multline}\label{eq:Ui_p}
  \frac{1}{P}U_{i} =  \frac{1}{P}
  \sum_{j=1}^{P}\1(\underline{X}(i)>\overline{X}(j)) - \1(\overline{X}(i)<\underline{X}(j))\\ 
  = F_{P}(\underline{X}(i)^{-}) - \{1-G_{P}(\overline{X}(i))\}
  = F_{P}(\underline{X}(i)^{-}) - \overline{G}_{P}(\overline{X}(i))\; ,
\end{multline}
where $\overline{G}_{P}(\cdot) = 1 - G_{P}(\cdot)$. Then, using the 
Glivenko-Cantelli Theorem \cite[Theorem 19.1]{vandervaart:1998}, we
get, as $P$ tends to infinity, that
\begin{multline*}
  \frac{1}{P} \sum_{j=1}^{P}\left(\frac{1}{P}U_{j}\right)^{2}\\
     =\frac{1}{P} \sum_{j=1}^{P} F_{P}(\underline{X}(i)^{-})^{2}
     -\frac{2}{P} \sum_{j=1}^{P}  F_{P}(\underline{X}(i)^{-}) \overline{G}_{p}(\overline{X}(i))
     +\frac{1}{P} \sum_{j=1}^{P} \overline{G}_{p}(\overline{X}(i))^{2} \\
      = \frac{1}{P} \sum_{j=1}^{P} F(\underline{X}(i)^{-})^{2}
     -\frac{2}{P} \sum_{j=1}^{P}  F(\underline{X}(i)^{-}) \overline{G}(\overline{X}(i))
     +\frac{1}{P} \sum_{j=1}^{P} \overline{G}(\overline{X}(i))^{2} +
     o_{p}(1) \; .
\end{multline*}
By the law of large numbers and our assumption in~(\ref{eq:cond1}), we obtain,
as $P$ tends to infinity, that
\begin{equation}
\label{eq:2}
  \frac{1}{P} \sum_{j=1}^{P}\left(\frac{1}{P}U_{j}\right)^{2}
   \inproba \PE[\{F(\underline{X}^{-})-\overline{G}(\overline{X})\}^{2}] >0\; .
\end{equation}
%
%
Using (\ref{eq:Ui_p}), $P^{-1}|U_{i}|\leq 2$, $i=1,\dotsc,P$, and thus
\begin{multline*}
 |Y_i|= \frac{|U_{i}|}{\sqrt{\sum_{j=1}^{P}U^{2}_{j}}} 
=\inv{\sqrt{P}}\frac{P^{-1}|U_{i}|}{\sqrt{P^{-1}\sum_{j=1}^{P}
     (P^{-1}U_{j})^{2}}}\\
 \leq \inv{\sqrt{P}}\frac{2}{\sqrt{P^{-1}\sum_{j=1}^{P}
     (P^{-1}U_{j})^{2}}}\; , \; i=1,\dots,P\; .
\end{multline*}
By (\ref{eq:2}), $Y_i$ satisfies the third condition of (\ref{eq:cond_billingsley}), which completes the proof.
\bibliographystyle{apalike}      

\small{
\bibliography{toprank}   
}

\end{document}